\documentclass[12pt,preprint]{aastex}

\newcommand{\be}{\begin{equation}}
\newcommand{\ee}{\end{equation}}
\newcommand{\erf}{ {\rm Erf} }  
 
\def\lta{\,\raise 0.3 ex\hbox{$ < $}\kern -0.75 em
 \lower 0.7 ex\hbox{$\sim$}\,}
\def\gta{\,\raise 0.3 ex\hbox{$ > $}\kern -0.75 em
 \lower 0.7 ex\hbox{$\sim$}\,} 

\newcommand{\fsuper}{ {\cal F}_{SN} }
\newcommand{\mbar}{ \langle m \rangle } 
\newcommand{\nmax}{{ N_{\rm max} }} 
\newcommand{\sigbar}{\langle\sigma\rangle}
\newcommand{\exstar}{\langle{M[A_k]}\rangle_\ast} 
\newcommand{\exstarsqu}{\langle{M[A_k]^2}\rangle_\ast} 
\newcommand{\exyieldn}{\langle{M[A_k;N]}\rangle}
\newcommand{\yieldn}{M[A_k;N]}
\newcommand{\yieldm}{M[A_k;m]}
\newcommand{\yieldj}{M[A_k;m_j]}
\newcommand{\halflife}{ \tau_{1/2}} 
\newcommand{\mmax}{m_\infty} 
\newcommand{\ionize}{ \omega_I } 
\newcommand{\taufree}{\tau_{\rm ff}} 
\newcommand{\sfe}{\epsilon_{\rm sfe}} 
\newcommand{\aconst}{ {\rm A}_{\rm Y}} 

\begin{document} 

\title{DISTRIBUTIONS OF SHORT-LIVED RADIOACTIVE NUCLEI \\
PRODUCED BY YOUNG EMBEDDED STAR CLUSTERS} 

\author{Fred C. Adams$^{1,2}$, Marco Fatuzzo$^3$, and Lisa Holden$^4$}  

\affil{$^1$Physics Department, University of Michigan, Ann Arbor, MI 48109} 

\affil{$^2$Astronomy Department, University of Michigan, Ann Arbor, MI 48109} 

\affil{$^3$Physics Department, Xavier University, Cincinatti, OH 45255} 

\affil{$^4$Department of Mathematics, Northern Kentucky University, Highland Heights, KY 41099} 

\begin{abstract}

Most star formation in the Galaxy takes place in clusters, where the
most massive members can affect the properties of other constituent
solar systems. This paper considers how clusters influence star
formation and forming planetary systems through nuclear enrichment
from supernova explosions, where massive stars deliver short-lived
radioactive nuclei (SLRs) to their local environment. The decay of
these nuclei leads to both heating and ionization, and thereby affects
disk evolution, disk chemistry, and the accompanying process of planet
formation. Nuclear enrichment can take place on two spatial scales:
[1] Within the cluster itself ($\ell\sim1$pc), the SLRs are delivered
to the circumstellar disks associated with other cluster members. [2]
On the next larger scale ($\ell\sim2-10$pc), SLRs are injected into
the background molecular cloud; these nuclei provide heating and
ionization to nearby star-forming regions, and to the next generation
of disks. For the first scenario, we construct the expected
distributions of radioactive enrichment levels provided by embedded
clusters. Clusters can account for the SLR mass fractions inferred for
the early Solar Nebula, but typical SLR abundances are lower by a
factor of $\sim10$. For the second scenario, we find that distributed
enrichment of SLRs in molecular clouds leads to comparable abundances.
For both the direct and distributed enrichment processes, the masses
of $^{26}$Al and $^{60}$Fe delivered to individual circumstellar disks
typically fall in the range $10-100pM_\odot$ (where
$1pM_\odot=10^{-12}M_\odot$).  The corresponding ionization rate due
to SLRs typically falls in the range $\zeta_{SLR}\sim1-5\times10^{-19}$
sec$^{-1}$. This ionization rate is smaller than that due to cosmic
rays, $\zeta_{CR}\sim10^{-17}$ sec$^{-1}$, but will be important in
regions where cosmic rays are attenuated (e.g., disk mid-planes). 

\end{abstract}


\section{Introduction} 
\label{sec:intro} 

Within our Galaxy, most star formation takes place within embedded
stellar clusters, where these systems display a wide range of sizes
and other properties (e.g, \citealt{ladalada,porras,allen}). These
background environments can influence the evolution and properties of
planetary systems forming within them through a variety of processes
\citep{hesterdesch,looney,levison,adams2010,pfalzner}, including
dynamical scattering by other stellar members
\citep{al2001,adams2006,malmberg}, evaporation of circumstellar disks
by radiation from massive stars \citep{storzer,adams2004}, accretion
of cluster gas onto the disks \citep{throop}, and the injection of
short-lived radioactive nuclei (hereafter SLRs) into circumstellar
disks and/or collapsing regions \citep{cameron,meyerclayton,ouellette}. 
This paper focuses on this latter issue of nuclear enrichment.  In
particular, we construct probability distributions for the abundances
of SLRs that are expected to be delivered to forming solar systems in
cluster environments, as well as to the larger, surrounding regions 
in molecular clouds. 

Previous work regarding nuclear enrichment has focused on two related
but somewhat different goals. The first approach considers the largest
spatial scales, where a great deal of work has been carried out to
estimate the steady state production rates and abundances of SLRs in
the Galaxy \citep{timmesa,timmesb}. For example, gamma-ray emission
due to the radioactive decay of $^{26}$Al has been observed for the
1808.65 keV line \citep{diehl,smith2003}. Since the Galaxy does not
appreciably attenuate such emission, it can be used to determine the
galactic inventory of $^{26}$Al; the abundances of other nuclear
species can be assessed in similar fashion.  Such measurements, in
conjunction with stellar evolution calculations that determine nuclear
yields \citep{ww1995,woosley2002,rauscher,lc2006}, can then be used to
estimate (or constrain) the Galactic star formation rate. In an
similar vein, gamma-ray observations have also measured the emission
due to decay of $^{60}$Fe. By comparing the ratio of line strengths,
one can test whether or not the inferred abundances of $^{60}$Fe and
$^{26}$Al are consistent with predictions of stellar nucleosynthesis
models \citep{prantzos}. The results are consistent at the
factor-of-two level, which is comparable to the uncertainties in these
quantities.

On smaller scales, our own Solar System was apparently enriched in
SLRs during its early formative stages. A great deal of previous work
has been carried out to explain the nuclear abundance patterns deduced
from meteoritic studies. Evidence for the enrichment of $^{26}$Al is
well-established, and the proposal that such enrichment arises from
nearby supernovae dates back to \cite{cameron77}, or even further.
Although $^{26}$Al is readily produced in supernovae, it can also be
synthesized through spallation.  More recently, meteoritic evidence
for live $^{60}$Fe in the early solar system has been reported
\citep{tachibana} and bolsters the case for supernova enrichment of
SLRs (because $^{60}$Fe can only be produced via stellar
nucleosynthesis). In contrast to the case of $^{26}$Al, however, the
evidence for $^{60}$Fe is controversial and elusive \citep{moynier}.
For example, the inferred abundances can be biased for low count 
rates, such as those found for $^{60}$Fe \citep{telus}. 
On the other hand, a recent study \citep{mishra} concludes 
that the two SLRs $^{26}$Al and $^{60}$Fe were co-injected into 
the early solar system from same stellar source. 
In any case, for the purposes of this paper, we consider the two
nuclear species to be on a nearly equal footing. As discussed below,
stellar evolution models predict comparable abundances for $^{26}$Al
and $^{60}$Fe, and both species are expected to make substantial
contributions to ionization and heating within enriched regions.

The apparent need for nuclear enrichment of the early Solar System
places constraints on the birth environment of the solar system
\citep{al2001,hesterdesch,williams,gounelle2008,gounelle2009,
  adams2010,pfalzner,dauphas}.  The requirement that the Sun is born
near a high mass progenitor favors the scenario where the Sun is
formed within a large cluster, but such environments can also lead to
disruption through dynamical scattering interactions
\citep{adams2006,malmberg,dukes} and intense radiation fields
\citep{fa2008,holden,thompson}. For the nuclear enrichment of our
Solar System, these previous studies (along with many others,
including \citealt{looney,williamsalt,parker}) suggest that the
probability of enrichment is low, perhaps 1--10\%, where the estimated
value depends on how the accounting is done.  On the other hand,
observations of accreted extrasolar asteroids in white dwarf
atmospheres indicate that the elevated levels of $^{26}$Al inferred
for the formation of own Solar System are not abnormal 
\citep{jura2013}.

Building upon the aforementioned previous work, this paper considers
nuclear enrichment on an intermediate scale. Instead of focusing on
the enrichment of our own Solar System, we consider the general
problem of determining the distribution of enrichment levels that are
expected for the whole population of forming stars within a cluster.
We also consider nuclear enrichment on the next larger scale of the
molecular cloud, but do not focus on the scale of the Galaxy.
Nonetheless, this work informs the larger scale picture, as the
results must be consistent with galaxy-wide estimates for nuclear
production and the accompanying star formation rate. On the scale of
our own Solar System, this work constrains the probability of
attaining the levels of nuclear enrichment inferred for our own solar
nebula.

The decay of SLRs is important for star and planet formation for two
related reasons: ionization and heating.  The nuclear decay products
are highly energetic, with $E \sim 1$ MeV. As these particles, mainly
photons and positrons, interact with their immediate environment,
their energy cascades down to photons of lower energy. After the
cascade proceeds to sufficiently low energy, the remaining photons
interact with hydrogen molecules (and/or atoms), often resulting in
ionization. Within circumstellar disks, ionization plays an important
role in setting the thermal and chemical properties of the material;
for example, the ionization state is crucial for the development of
the magneto-rotational instability (MRI), the process that helps drive
disk accretion \citep{balbushawley}. On the larger scales of molecular
clouds, and their constituent cloud cores, ionization determines, in
part, the coupling between the gas (which is mostly neutral) and the
magnetic field \citep{shubook}. 

This paper is organized as follows: In Section \ref{sec:environments}
we review the stellar initial mass function (IMF), as well as the
distributions of cluster masses and other properties.  The predicted
yields of SLRs are discussed in Section \ref{sec:slr}, including both
the nuclear yields produced by supernovae and the corresponding yields
expected per star (obtained by averaging over the stellar IMF). The
nuclear yields per cluster are considered next, in Section
\ref{sec:clustyield}, where we also determine expectation values for
the SLR yields per cluster (averaging over the cluster distribution)
and the yields expected for an average star in a cluster.  The
distributions of the SLR abundances delivered to individual solar
systems are then constructed in Section \ref{sec:systemyield}.
Global considerations are discussed in Section \ref{sec:global},
including the galaxy-wide star formation rate and molecular cloud
inventories of SLRs. Finally, we conclude, in Section
\ref{sec:conclude}, with a summary of our results and a discussion of
their implications.

\section{Stellar and Cluster Mass Distributions} 
\label{sec:environments} 

\subsection{Initial Mass Function for Stars} 

The distribution of SLRs will depend on the stellar initial mass
function (IMF). This distribution has been studied intensively (see
\citealt{kroupa,chabrier} for recent reviews). In its most basic form,
the stellar IMF has the form of a log-normal distribution with a
power-law tail on the high-mass end and perhaps another tail at the
low-mass end \citep{af1996}.  In any case, the distribution is heavily
weighted toward stars of low mass (see also \citealt{salpeter,scalo}).
In this work, however, we are interested in the high-mass end of the
stellar IMF, as only stars with mass above the threshold 
$M_\ast \gta 8 M_\odot$ contribute to the supply of SLRs through
supernova explosions.

For the present application, one useful way to parameterize the
stellar IMF is to define $\fsuper$ to be the fraction of the (initial)
stellar population with masses greater than the minimum mass
($M_\ast=8M_\odot$) required for a star to ends its life with a
supernova explosion.  Current observations indicate that
$\fsuper\approx0.005$, although this value remains uncertain (e.g.,
see the models advocated by \citealt{af1996,kroupa,chabrier}). The
mass distribution for massive stars can be written in power-law form 
\be
{dN_\ast \over dm} = \fsuper {\gamma \over 8} 
\left( {m \over 8} \right)^{-(\gamma+1)} \, , 
\label{imf} 
\ee
where $m=M_\ast/(1M_\odot)$ is the mass in Solar units and the
canonical value of the index $\gamma \approx$ 1.35
\citep{salpeter}. In spite of the large number of studies of the
stellar IMF, in many different environments, the power-law form
(\ref{imf}) for the high-mass end of the distribution remains
robust. However, the value of the index $\gamma$ appears to have
significant scatter from region to region (e.g., \citealt{scalo}),
such that $\gamma$ is evenly distributed within the range $\gamma$ =
$1.5\pm0.5$. For the sake of definiteness, we use $\gamma$ = 1.5 as
the default value to characterize the high-mass end of the IMF, but
allow the index to vary. 

As written, the probability distribution of equation (\ref{imf}) is
normalized so that
\be
\int_8^\infty {dN_\ast \over dm} dm = \fsuper\,. 
\label{norm} 
\ee 
As a result, the distribution is normalized to unity for the entire
mass distribution (not just the high mass end), where this statement
holds in the absence of a maximum stellar mass.  In practice, one
expects the stellar IMF to have a maximum stellar mass $\mmax$,
although the value of $\mmax$ remains uncertain.  In order of
magnitude, however, both theory and observations suggest that
$\mmax \approx 120$ is a good approximation. If we include this
upper mass limit in the integral of equation (\ref{norm}), we obtain 
a correction factor $[1-(8/\mmax)^{\gamma}]\approx0.983$ in the
normalization. Note that the relative size of this correction factor
is much smaller than the uncertainties in the other parameters that
specify the IMF. As a result, we ignore this correction for the
remainder of this work.

To complete the specification of the stellar IMF we also need the 
average stellar mass, i.e., 
\be
\mbar \equiv \int {dN_\ast \over dm} m dm \,. 
\ee
For example, since stellar clusters are often described by their
stellar membership size $N$ and/or their mass in stars $M$, we often
need to convert between the two (where $M=N\mbar$).  Similarly, star
formation rates can be specified in terms of `stars per unit time' or
`solar masses per unit time', with the conversion factor $\mbar$.

To summarize, this paper characterizes the stellar IMF in terms of the
three parameters $(\gamma,\fsuper,\mbar)$. Since the stellar IMF is
steeply declining for all possible choices of $\gamma$, most supernova
progenitors have masses in the range $m=8-25$. As a result, the most
important parameter that determines radioactive yields is the fraction
of stars $\fsuper$ that are above the mass threshold for supernovae.

\subsection{Mass Distribution and Properties of Embedded Star Clusters} 

Stars form within embedded clusters, where the number of members in
these systems spans a wide range (e.g., see the reviews of
\citealt{ladalada,porras,allen}). Unfortunately, current observations
are not sufficiently complete to specify the distribution of cluster
membership sizes $N$. Studies of embedded clusters in the Solar
neighborhood indicate that the distribution of cluster sizes $N$ is
close to a power-law, so that 
\be
{d N_C \over dN} = {C_N \over N^a} \, , 
\label{clusterdist} 
\ee
where $C_N$ is a normalization constant and where the index
$a\approx2$.  Here, $N_C$ is the number of clusters and $N$ is the
number of stellar members in the clusters. Studies that consider more
distant clusters also find power-law distributions, again with
$a\approx2$ (e.g., \citealt{bruce1997,kroupaboil,whitmore}). The
simplest possibility is that the cluster distribution has the same
power-law form, and the same normalization constant, over the entire
range of cluster sizes $N=1-10^6$. For sake of definiteness, we use a
single power-law in the present analysis (but one should keep in mind
that the power-law distribution observed for small clusters nearby and
that observed for large clusters at large distances have not been
shown to match up).  Further, we adopt a benchmark value of the index
$a$ = 2. In this case, the constant $C_N \approx 1$.

A related quantitiy is the probability $P_\ast (N)$ of a star being
born within a cluster of membership size $N$.  This probability
distribution is obtaining by multiplying $dN_C/dN$ by another factor 
of $N$ and hence is given by 
\be
P_\ast(N) = {C_P \over N} \, , \qquad {\rm with} \qquad 
C_P = {1 \over \log[\nmax]} \, , 
\label{probdist} 
\ee 
where we have taken $a=2$ and assume that clusters range from $N=1$ to
$N=\nmax$.  We expect $\nmax\approx10^6$ and hence $C_P \approx 1/(6
\log 10)$.  The cummulative probability ${\cal P}$ for finding stars
in clusters with membership size $N$ or smaller is thus given by
\be
{\cal P} (N) = C_P \log [N] = {\log [N] \over \log [\nmax]} \, . 
\ee
As a result, the probability of finding stars in various sized 
clusters is evenly distributed in a logarthmic sense.  

Next we must specify the radial extent of the cluster. To a reasonable
degree of approximation, the cluster radius $R$ can be written as a
power-law function of the cluster membership size $N$, i.e., 
\be
R = R(N) = R_0 \left( {N \over N_0} \right)^{\alpha} \,.
\label{rvsn} 
\ee
For the clusters found in the solar neighborhood
\citep{ladalada,porras}, equation (\ref{rvsn}) works well with $R_0$ =
1 pc, $N_0$ = 300, and $\alpha$ = 1/2. The cluster sample in the solar
neighborhood is limited to the lower end of the cluster membership
size distribution -- the sample is not large enough to include the
largest clusters. If one applies equation (\ref{rvsn}) to the entire
cluster sample (the entire range of $N$), then the predicted radii are
too large for high-mass clusters. We obtain an adequate fit over the
entire cluster size range, $10 \le N \le \nmax$, by using index
$\alpha$ = 1/3.

Finally, we need to specify the spatial distribution of stars within
the cluster. For simplicity, we assume that the stars in the cluster
follow a simple power-law distribution of density. Numerical (N-body)
simulations of early cluster dynamics \citep{adams2006} show that this
assumption is reasonable and indicates that the power-law index of the
density distribution falls in the range $1\le{p}\le2$.  As a result, 
the probability distribution for the radial distance (at a given time,
including the time of the supernova explosion) is given by
\be
{dP \over dr} = {4 \pi r^2 \over N} n_\ast(r) = 
{3-p \over R} \left( {r \over R} \right)^{2-p} \,,
\label{dpdr} 
\ee
where $R$ is the cluster radius (which varies with cluster 
membership size $N$ -- see equation [\ref{rvsn}]).  

\section{Production of Short-lived Radioactive Nuclei}  
\label{sec:slr} 

\subsection{Synthesis of SLRs in Supernovae} 
\label{sec:production} 

\begin{figure} 
\figurenum{1} 
{\centerline{\epsscale{0.90} \plotone{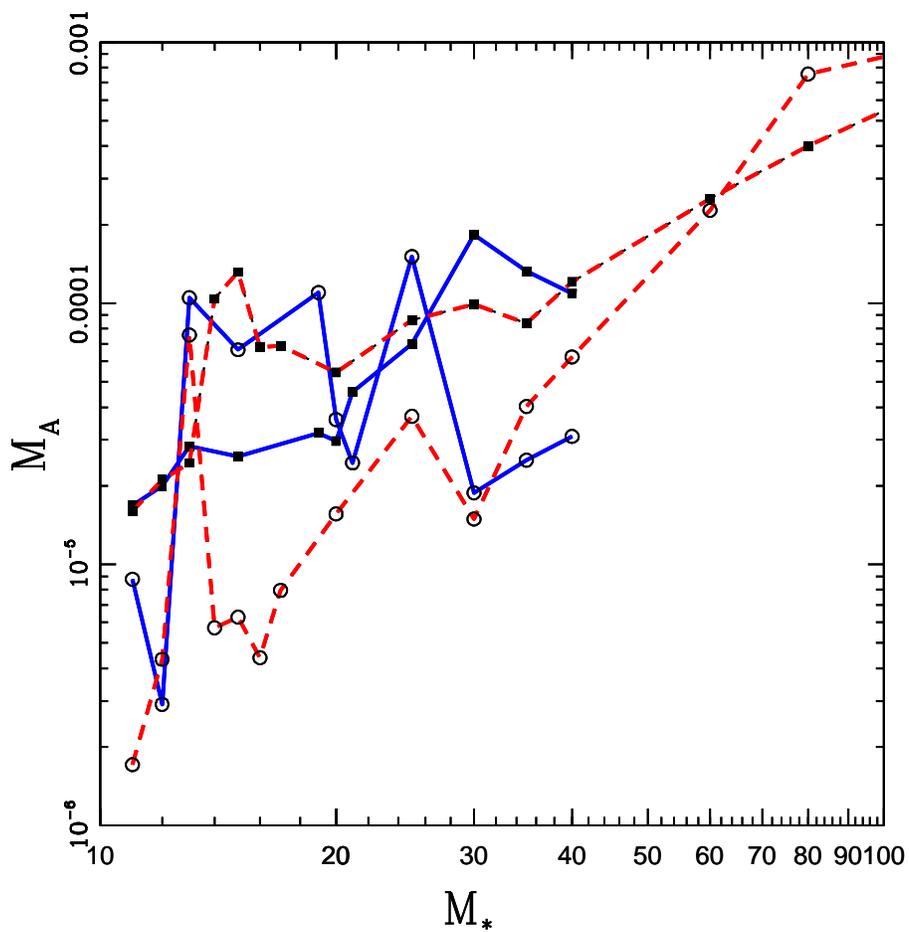} } } 
\figcaption{Yield $M_A$ of short-lived radioactive nuclei as a function 
of progenitor mass (both quantities are given in units of $M_\odot$).
The curves marked by solid squares indicate the yields for $^{26}$Al;
the curves marked by open circles indicate yields for $^{60}$Fe.
Results are presented for two different sets of stellar evolution
calculations: Yields from the WW models are shown as blue solid curves
\citep{ww1995,rauscher}, whereas yields from the LC models are shown
as red dashed curves \citep{lc2006}. } 
\label{fig:yieldalfe} 
\end{figure}

\begin{figure} 
\figurenum{2} 
{\centerline{\epsscale{0.90} \plotone{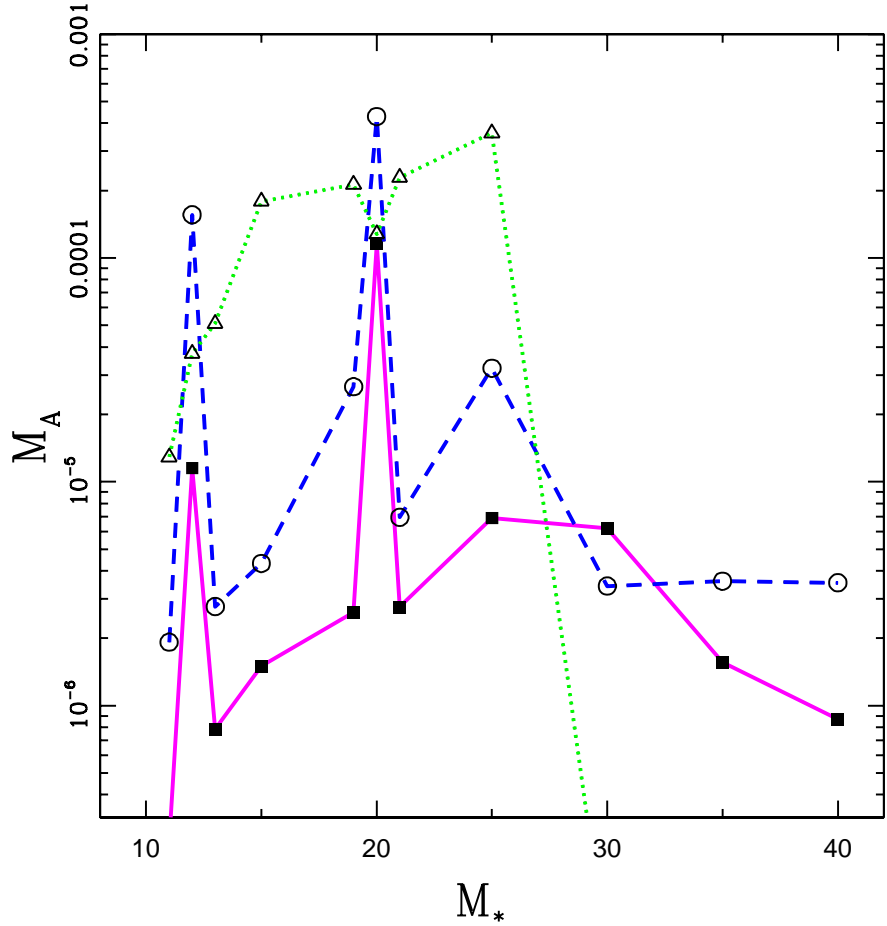} } } 
\figcaption{Yield $M_A$ of short-lived radioactive nuclei as a function 
of progenitor star mass (in units of $M_\odot$).  Solid magenta curve
marked by solid squares indicates the yields for $^{36}$Cl; the dashed
blue curve marked by open circles provides the yield for $^{41}$Ca;
the dotted green curve marked by open triangles provides the yields
for $^{53}$Mn. Results are taken from the WW stellar evolution
calculations \citep{ww1995,woosley2002,rauscher}. }
\label{fig:yieldclca} 
\end{figure}

Supernovae produce a wide variety of radioactive nuclei. In this
treatment, we consider only the production and distributions of the
five most important species of SLRs, namely $^{26}$Al, $^{36}$Cl,
$^{41}$Ca, $^{53}$Mn, and $^{60}$Fe.  These species all have
half-lives shorter than 10 Myr, and relatively large abundances; these
properties, in turn, make them useful for constraining the early
history of our Solar System \citep{cameron,meyerclayton,ouellette}.
The abundances of these SLRs are shown as a function of stellar mass
in Figures \ref{fig:yieldalfe} and \ref{fig:yieldclca}, with the
yields taken from the calculations of \cite{ww1995}, \cite{rauscher},
and \cite{lc2006}. 

For purposes of finding nuclear enrichment levels for typical solar
systems, we often further limit our focus to the two species $^{26}$Al
and $^{60}$Fe, because they provide the largest contribution to the
ionization and heating rates. Figure \ref{fig:yieldalfe} shows the
yields for these two SLRs, where we include results from two different
sets of stellar nucleosynthesis calculations. The first group
\citep{ww1995} considers the range of progenitor masses $M_\ast=11-40
M_\odot$, where these results have been updated \citep{rauscher} for
the more limited range in stellar masses $M_\ast=15-25M_\odot$ (see
also \citealt{timmesa,timmesb,woosley2002}). For the yields shown in
Figure \ref{fig:yieldalfe}, we use the updated yields for the range of
masses where they are available; hereafter results from this set of
papers are denoted as WW.  The second set of results \citep{lc2006},
hereafter LC, considers a wider range of masses $M_\ast=11-120M_\odot$. 
As shown in the Figure, the two different sets of nucleosynthesis
calculations are not in perfect agreement. Although the predicted
yields used here only vary by a factor of $\sim2$, and this level of
uncertainty is often quoted, we note that even larger variations are
possible. The predicted abundances of both $^{26}$Al and $^{60}$Fe are
extremely sensitive to variations in the triple-$\alpha$ reaction
\citep{tur2010}, and the corresponding reaction rates are not
precisely known. As a result, the uncertainties in the yields for SLRs
can be {\it larger} than a factor of two. Finally, we note that the
yields for the WW models are only given up to $m=40$. Here we 
extrapolate to larger values assuming $M_A$ = {\sl constant} for 
$m>40$; although stars more massive than this threshold are rare, 
this choice provides another source of uncertainty.  

For completeness, we show the expected yields of the isotopes
$^{36}$Cl, $^{41}$Ca, and $^{53}$Mn in Figure \ref{fig:yieldclca}
(where these results are taken from the WW group).  The abundance of
$^{36}$Cl is lower than those of $^{26}$Al and $^{60}$Fe by roughly an
order of magnitude, so that its contribution to the ionization rate is
correspondingly smaller. On the other hand, while $^{41}$Ca and
$^{53}$Mn have relatively large abundances, they both decay via
electron capture, and do not (immediately) emit ionizing energy.

We note that this analysis assumes that all of the SLRs produced
by supernova explosions are actually ejected outward and become
available for enrichment. In practice, however, calculations of
supernovae remain challenging \citep{mosta}, so that the degree of
mixing and fallback of synthesized material is not completely known.
Indeed, some cosmochemical models suggest some fallback is necessary
to explain the particular isotopic abundances observed in the solar
system (e.g., \citealt{liu}). A significant amount of fallback would
lower the SLR abundances determined in this paper. 

\subsection{Nuclear Yields per Star} 

For a cluster of given membership size $N$, and for a given stellar
IMF, we want to find the abundance distribution for each type of
radioactive species (here we denote an arbitrary radionuclide as $A_k$).
As outlined below, we can determine these abundance distributions via
sampling. To leading order, however, the distributions can be
characterized by their expectation values and widths (or variance).
At higher order, however, the distributions show significant
departures from gaussians.

To start, we define the yield weighted by the stellar IMF. More
specifically, for a given IMF and radionuclide $A_k$, we define the
expectation value per star $\exstar$ of the yield 
to be  
\be
\exstar \equiv 
\int_{m_{min}}^{\mmax} \yieldm {dN_\star \over dm} dm \, ,
\label{radiomean} 
\ee 
where $\yieldm$ is the yield of species $A_k$ for a progenitor star 
of mass $m$.  The upper limit of integration is set by the maximum
stellar mass, which is taken here to be $\mmax=120$; since the
stellar IMF is a steeply decreasing function of stellar mass, and
since the expected yields are not steeply increasing with mass, most
of the support for the integral in equation (\ref{radiomean}) occurs
for the smaller masses, so that results are not overly sensitive to
the upper limit. The lower limit of integration is set by the minimum
mass required for a star to explode and thereby provide nuclear
enrichment; this criterion thus implies that $m_{min}\sim8$. However,
the main-sequence lifetime of stars with $m\sim8$, about 20 Myr, is
longer than the expected time for which circumstellar disks retain
their gas (3 -- 10 Myr; \citealt{jesus}), and is longer than the
lifetime of most embedded clusters (3 -- 10 Myr;
\citealt{allen,guter2009}).  More recent studies based on 
{\it Spitzer} observations \citep{cieza}, and SCUBA-2 surveys
\citep{williams2013}, suggest that the lifetime of circumstellar disks
could be even shorter, 1 -- 3 Myr.  As a result, although we use the
full range of stellar masses (with $m_{min}=8$) to compute expectation
values, these results can be subject to a reduction factor because of
constraints on stellar lifetimes.  Finally, we note that the
expectation value defined via equation (\ref{radiomean}) is normalized
so that it provides the expected radioactive yield {\it per star}.
Because only the massive stars contribute to the yields, this
expectation value per star is much smaller than the radioactive yield
per supernova; these yields are smaller by the factor $\fsuper$.  For
the sake of definiteness, we use $\fsuper\approx0.005$ as a standard
benchmark value; the nuclear yields can be scaled upward, or downward,
for alternate values of $\fsuper$.

\begin{figure} 
\figurenum{3} 
{\centerline{\epsscale{0.90} \plotone{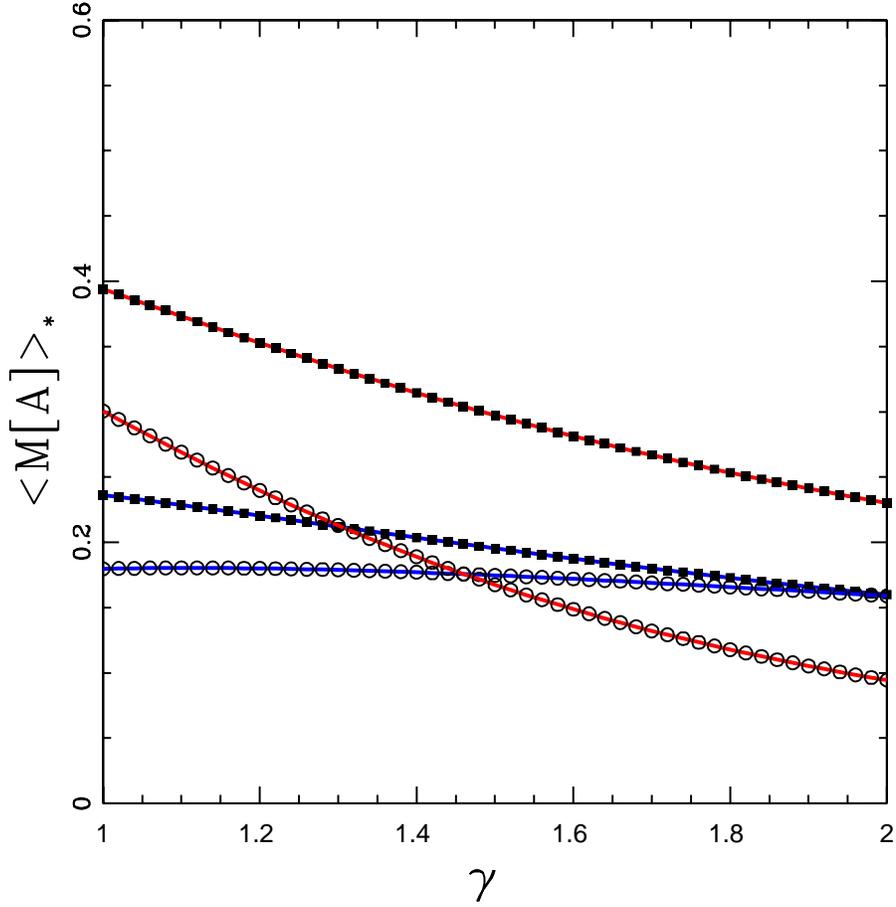} } } 
\figcaption{Radioactive yields per star for $^{26}$Al and $^{60}$Fe 
versus index $\gamma$ of the stellar IMF. The yields, which are given
in units of $\mu M_\odot = 10^{-6} M_\odot$, are proportional to the
fraction of stars above the supenova mass threshold, taken here to be
$\fsuper=0.005$.  Results are shown for the two different sets of
stellar evolution calculations, WW (blue solid curves), and LC (red
dashed curves). The curves presenting yields for $^{26}$Al are marked
by solid square symbols, whereas the corresponding curves for
$^{60}$Fe are marked by open circles. }
\label{fig:yvgamma} 
\end{figure}

For the two SLRs of greatest interest, $^{26}$Al and $^{60}$Fe, Figure
\ref{fig:yvgamma} presents the expectation value for the yield per
star.  These yields are plotted here as a function of the index
$\gamma$ of the stellar initial mass function. Note that the yields
are given in units of ``microsuns'' $\mu M_\odot$ (where $1 \mu
M_\odot = 10^{-6} M_\odot$). For the WW models of stellar evolution,
the nuclear yields vary slowly with index $\gamma$ and the values for
$^{26}$Al and $^{60}$Fe are nearly equal (see the blue solid curves in
Figure \ref{fig:yvgamma}). In contrast, for the stellar models of LC,
the nuclear yields vary by a factor of $\sim3$ over the possible range
of the index $\gamma$ (see the red dashed curves in the Figure). In
addition, the yield for $^{26}$Al is significantly larger than that
for $^{60}$Fe. The variations shown in Figure \ref{fig:yvgamma} provide
a measure of the uncertainty in the yields, due to possible variations
in the index of the IMF and/or uncertainties in the stellar
nucleosynthesis calculations. 

\begin{figure} 
\figurenum{4} 
{\centerline{\epsscale{0.90} \plotone{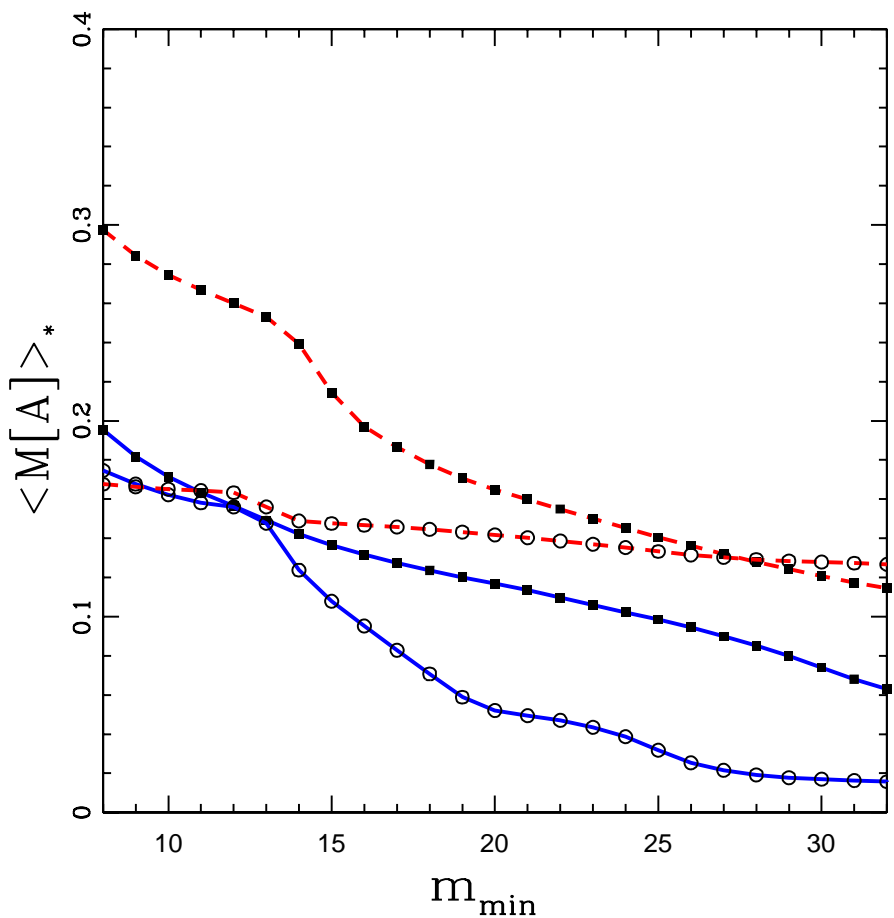} } } 
\figcaption{Radioactive yields per star for $^{26}$Al and $^{60}$Fe 
versus minimum mass of progenitor star included in the distribution.
The yields are given in units of $\mu M_\odot = 10^{-6} M_\odot$ and
the index of the stellar IMF $\gamma$ = 1.5.  Results are shown for
the two different sets of stellar evolution calculations, WW (blue
solid curves), and LC (red dashed curves). The curves presenting
yields for $^{26}$Al are marked by solid square symbols, whereas the
corresponding curves for $^{60}$Fe are marked by open circles. } 
\label{fig:yvmass} 
\end{figure}  

\begin{figure} 
\figurenum{5} 
{\centerline{\epsscale{0.90} \plotone{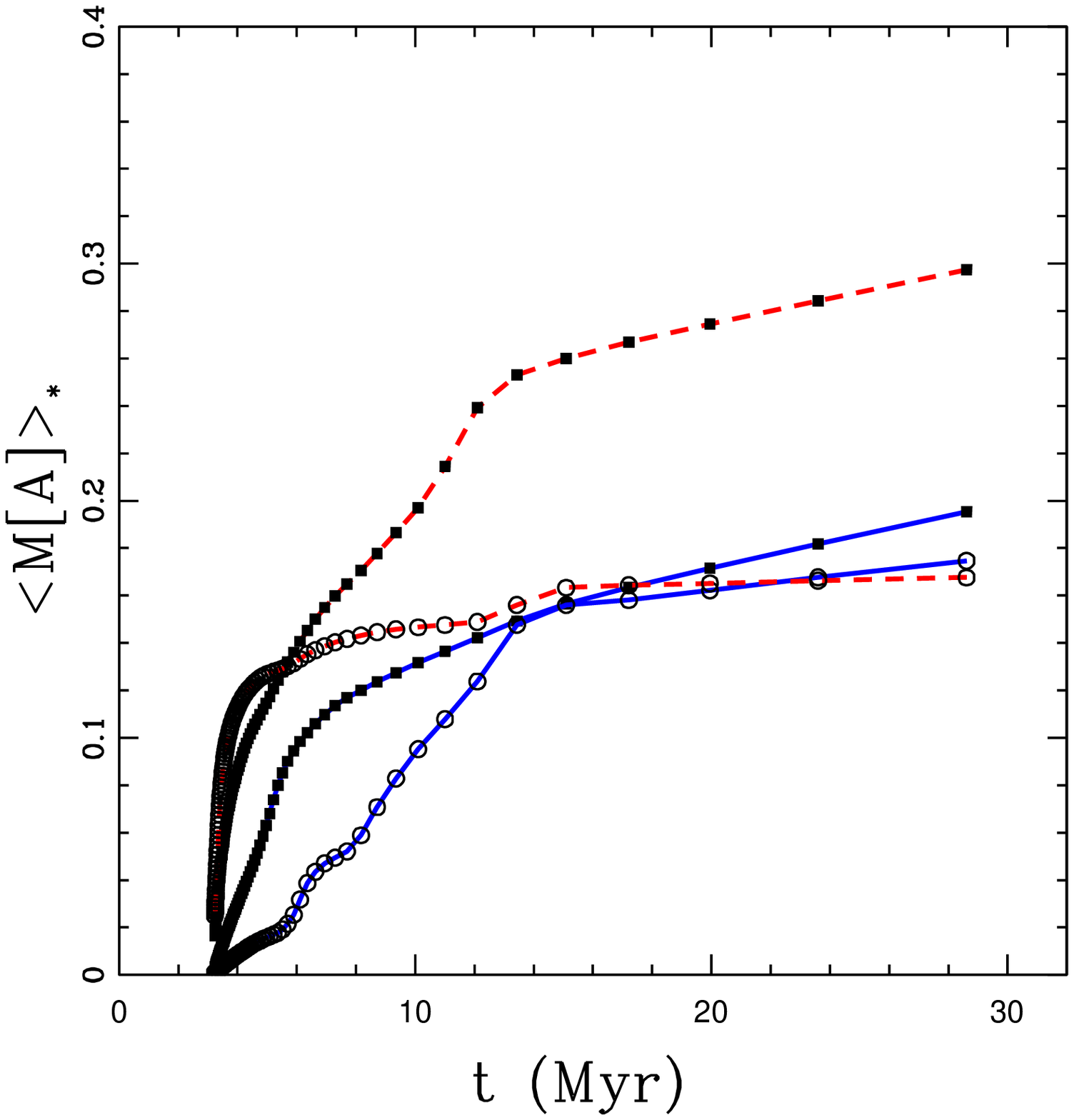} } } 
\figcaption{Radioactive yields per star for $^{26}$Al and $^{60}$Fe 
versus time (in Gyr). For a given time, only those stars that have
evolved enough to explode as supernovae are included in the integral
over the stellar mass distribution. The yields are given in units of
$\mu M_\odot = 10^{-6} M_\odot$ and the index of the stellar IMF
$\gamma$ = 1.5.  Results are shown for the two different sets of
stellar evolution calculations, WW (blue solid curves), and LC (red
dashed curves). The curves presenting yields for $^{26}$Al are marked
by solid square symbols, whereas the corresponding curves for
$^{60}$Fe are marked by open circles. }
\label{fig:yvtime} 
\end{figure}  

Another source of variation in the nuclear yields arises because the
lifetimes of massive stars vary with stellar mass, and these
timescales are comparable to the times over which both clusters and
disks remain intact. If we consider sufficiently short time scales for
disk and cluster evolution, we need to calculate the nuclear yields
produced by only those stars with the highest masses. To quantify this
effect, Figure \ref{fig:yvmass} presents the nuclear yields per star,
calculated using equation (\ref{radiomean}) with different values for
the minimum stellar mass $m_{min}$. As expected, the yields (per star)
decrease steadily with increasing minimum mass. Many previous studies
for nuclear enrichment of our Solar System use a progenitor with
$m=25$ as a standard value; a star with this mass spends 6.7 Myr
burning hydrogen and a total time of about 7.5 Myr before experiencing
core collapse \citep{woosley2002}.  Figure \ref{fig:yvmass} indicates
that the yields determined with $m_{min}=25$ are lower than those
obtained with the full spectrum of stellar masses by a factor of
$\sim2$ (where this factor was obtained by averaging over the four
curves shown in the Figure). The $^{26}$Al yields from the WW models
and the $^{60}$Fe yields from the LC models vary by less than this
factor, whereas the $^{26}$Al yields from the LC models and the
$^{60}$Fe yields from the WW models vary by a larger factor. 

We can also plot the nuclear yields as a function of time, as shown in
Figure \ref{fig:yvtime}, where only stars that have evolved far enough
to explode in time $t$ are included in the yield.  This figure is
essentially equivalent to Figure \ref{fig:yvmass}, where the minimum
mass $m_{min}$ is converted into the time required for a star of the
given mass to evolve and explode as a supernovae. Stars of the highest
mass ($m \sim 100$) have the smallest lifetimes $t \sim 3$ Myr; as a
result, the yields are zero for shorter times $t < 3$ Myr. The plot in
Figure \ref{fig:yvtime} extends out to about 28.6 Myr, which is the
time required for the star with the smallest mass to explode (among
those stars that are large enough to end their lives as supernovae).
For time scales that exceed this benchmark, all potential supernova
progenitors have time to evolve to completion, and the nuclear yields
reach their full asymptotic values. Of course, SLRs also decay during
such long time intervals (see, e.g., Section \ref{sec:time}). 
Although the results vary between the two stellar evolution models 
(WW and LC), and between the two types of SLRs under consideration
($^{26}$Al and $^{60}$Fe), all of the yields reach about half of their
asymptotic values by $t\sim10$ Myr.

\begin{figure} 
\figurenum{6} 
{\centerline{\epsscale{0.90} \plotone{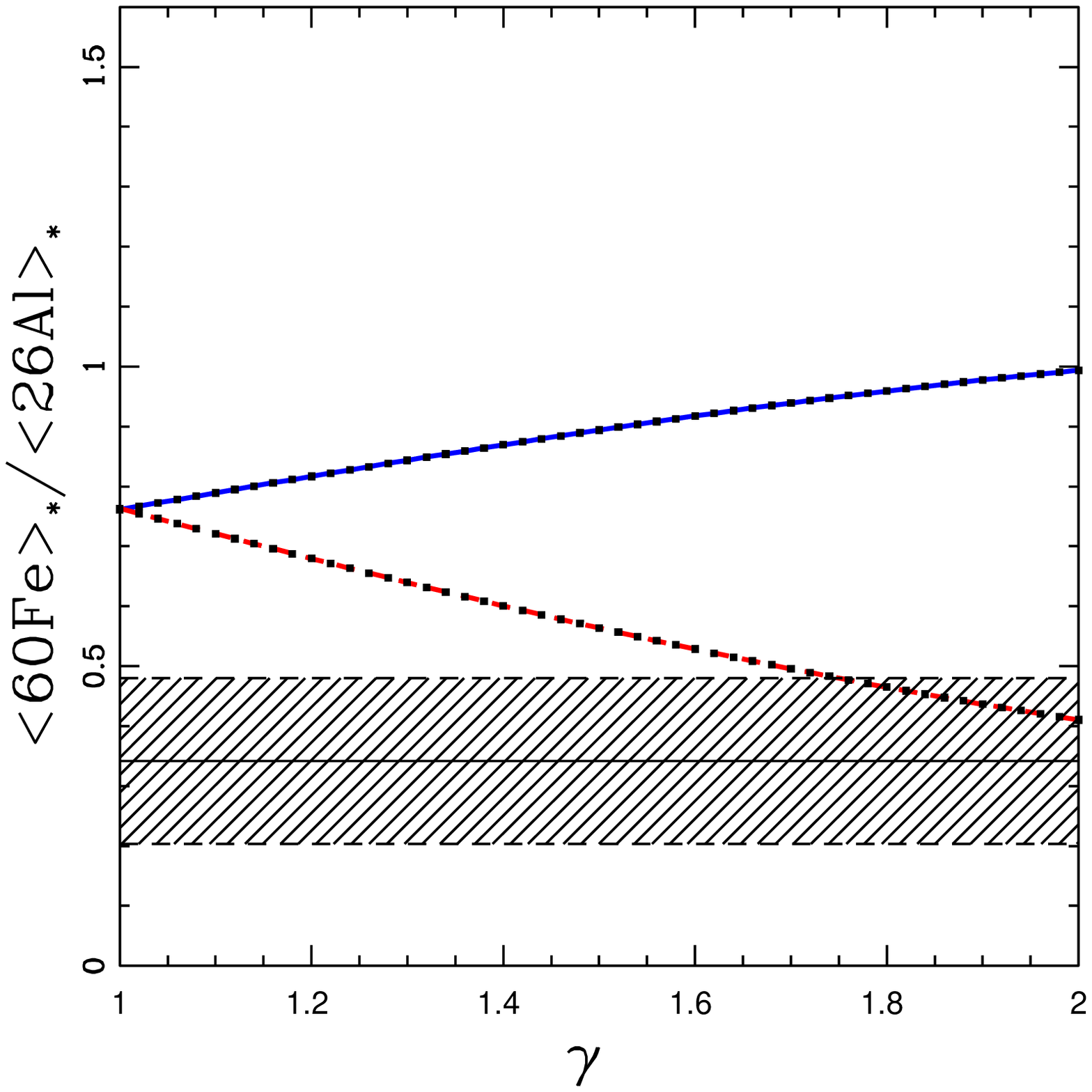} } } 
\figcaption{Ratio of radioactive yields per star for $^{60}$Fe and 
$^{26}$Al, plotted here as a function of the index $\gamma$ of the  
stellar IMF (where the yields are given in units of mass).  Results
are shown for the two sets of stellar evolution calculations, WW (blue
solid curve), and LC (red dashed curve).  Horizontal line depicts the
observed ratio \citep{diehl,diehlreview}, where the shaded region 
delineates the uncertainty in the measurement. } 
\label{fig:massratgam} 
\end{figure}  

\begin{figure} 
\figurenum{7} 
{\centerline{\epsscale{0.90} \plotone{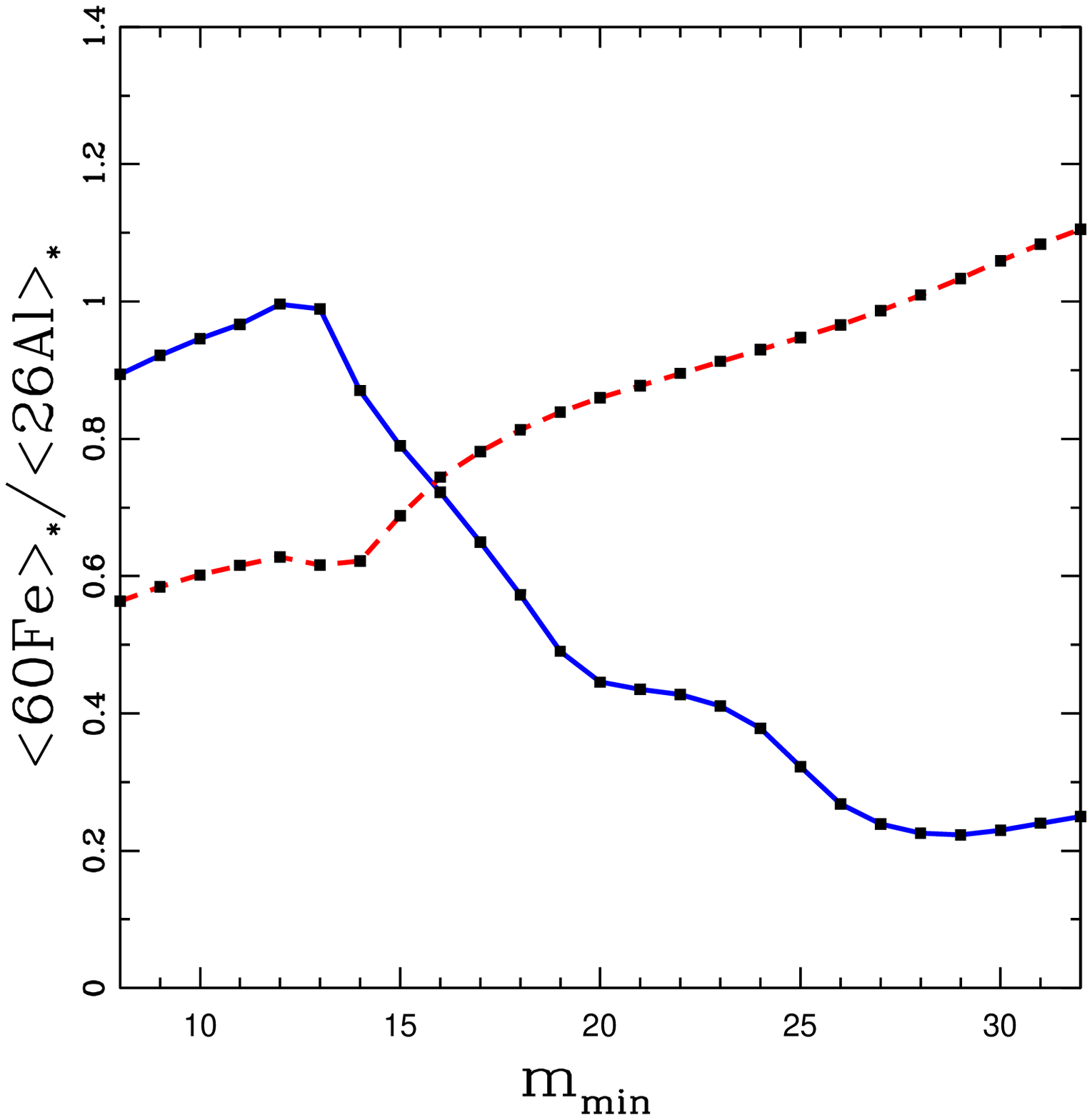} } } 
\figcaption{Ratio of radioactive yields per star for $^{60}$Fe and 
$^{26}$Al, plotted here as a function of the minimum mass of progenitor 
star included in the distribution. The yields are given in units of
mass, and the index of the stellar IMF $\gamma$ = 1.5.  Results are
shown for the two sets of stellar evolution calculations, WW (blue
solid curve), and LC (red dashed curve). }
\label{fig:massratio} 
\end{figure}  

A related quantity of interest is the ratio of the mass in $^{60}$Fe
produced to that of $^{26}$Al. Figure \ref{fig:massratgam} shows this
mass ratio as a function of the IMF index $\gamma$. Here, the yields
of both nuclear species are averaged over the stellar IMF, and then
the mass ratio is found (reversing the order of these operations would
produce a different result).  This mass ratio also depends on the
minimum mass included in the determination of the radioactive yields;
the resulting mass ratio is shown as a function of the minimum mass in
Figure \ref{fig:massratio}.  Whereas the overall yields must decline
with increasing minimum mass (Figure \ref{fig:yvmass}), the mass ratio
displays more complicated behavior. As the index $\gamma$ increases,
the stellar IMF is weighted more toward stars of lower masses; the
mass ratio for the WW models increases, whereas the mass ratio for the
LC models decreases.  Similarly, as the minimum mass increases, the
predicted mass ratio decreases for the WW models of stellar evolution,
but increases for the LC models.

\subsection{Time Evolution} 
\label{sec:time} 

The different species of radioactive nuclei decay at different rates,
so their relatative abundances vary with time. Since one important
implication of this work is the ionization provided by SLRs, we use
the ionization rate to illustate this time dependence. In addition,
both molecular clouds and circumstellar disks, the two environments of
interest, are composed primarily of H$_2$, so that we focus on the
ionization of molecular Hydrogen. Ionization rates for other species
can be scaled accordingly. For a given nuclear species, labeled by the
index `$k$', the ionization rate $\zeta_k$ per hydrogen molecule
\citep{umenakano} is given by the expression 
\be
\zeta_k = {E_k\over\ionize} \tau_k^{-1} X_k A_k^{-1} \exp[-t/\tau_k]\,,
\label{zetadef} 
\ee
where $E_k$ is the energy per decay ($\sim 1$MeV) of the given
nucleus, $\tau_k = \tau_{1/2}/\ln2$ is the decay time, $X_k$ is the
mass fraction, and $A_k$ is the atomic weight. The parameter
$\ionize\approx36$ eV is the average energy required for an energetic
particle to produce an electron-ion pair by passing through H$_2$ gas
\citep{umenakano}. 

The two most important SLRs are $^{26}$Al and $^{60}$Fe, which have
abundances that are roughly equal when averaged over the stellar IMF
(see Figure \ref{fig:yvgamma} for further detail).  Because of the
difference in half-lives, however, $^{26}$Al (with $\tau_{1/2}$ = 0.72
Myr; \citealt{rightmire,norris}) will dominate the ionization rate at
early times (measured from the time of the supernova explosion), and
$^{60}$Fe (with $\tau_{1/2}$ = 2.6 Myr; \citealt{rugel}), will
dominate at later times. The third species of possible interest,
$^{36}$Cl, has a smaller abundance and a shorter half-life
($\tau_{1/2}$ = 0.3 Myr; \citealt{ekfire}).  In addition, the net
energy per decay for $^{36}$Cl is only about 0.286 MeV, which is
appreciably smaller than that of $^{26}$Al (3.065 MeV) and $^{60}$Fe
(2.741 MeV); these values were obtained by averaging the decay energy
over the various channels, weighted by the branching ratios, by using
the data presented in Table 2 of \cite{umenakano}.

\begin{figure} 
\figurenum{8} 
{\centerline{\epsscale{0.90} \plotone{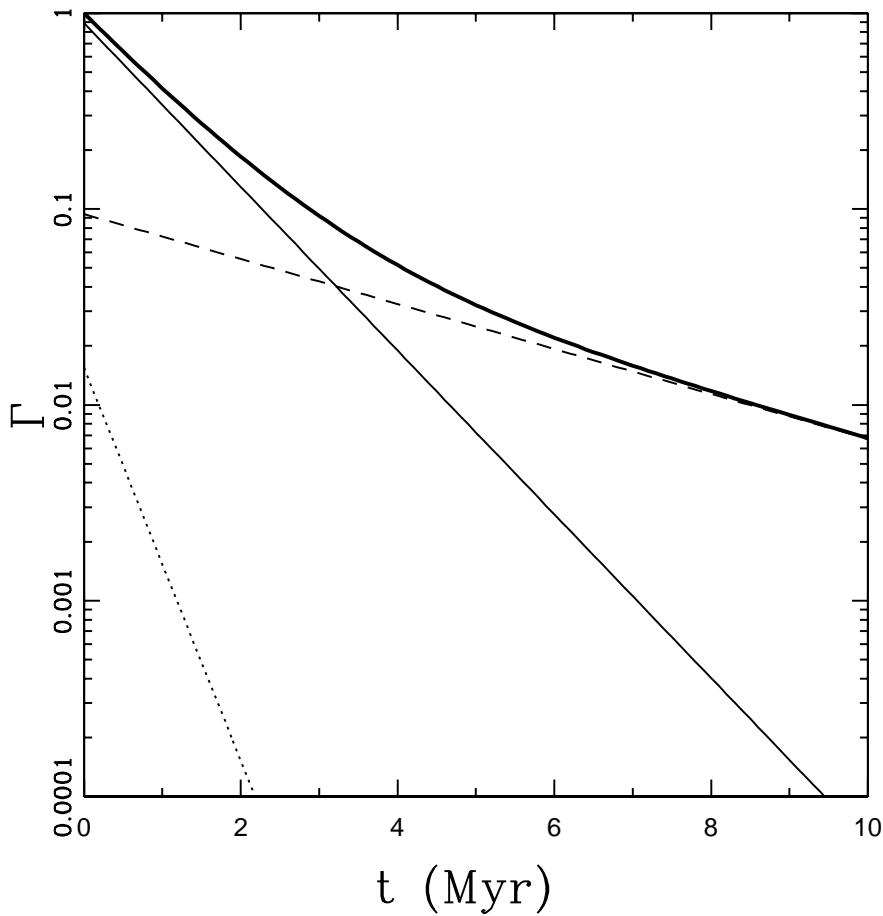} } } 
\figcaption{Contributions of the three most important nuclear species
to the SLR-induced ionization rate (using nuclear yields from the WW
stellar evolution models).  The ionization rate is normalized to unity
at $t=0$, the time of the supernova explosion. The relative abundances
are determined by using the yields per star, which have been averaged
over the stellar IMF. The curves correspond to $^{26}$Al (solid),
$^{36}$Cl (dotted), $^{60}$Fe (dashed), and the total (heavy solid). }
\label{fig:ionize} 
\end{figure}

The resulting time dependence of the ionization rate is illustrated in
Figure \ref{fig:ionize}. Here, we model the averaged expected behavior
by considering an effective ``supernova'' that produces the
IMF-averaged yields at $t=0$.  The resulting ionization rate is
normalized so that the sum of the contributions from the three SLRs
($^{26}$Al, $^{36}$Cl, and $^{60}$Fe) is unity at $t=0$.  The time
evolution of the ionization rate then shows the expected behavior: The
contribution of $^{36}$Cl is minimal and becomes less important with
time.  Ionization due to $^{26}$Al dominates at early times, whereas
that due to $^{60}$Fe dominates at later times, with the crossover
occurring at about $t\approx3.2$ Myr.

In order to determine the magnitude of the ionization rate, we need to
specify the amount of material that is mixed with the supernova ejecta.  
The radioactive yields (e.g., see Table 1) are given in terms of the
mass of SLRs per star (i.e., averaged over the stellar IMF). If we
consider the SLRs to be mixed with 1.0 $M_\odot$ of material, for
example, the ionization rate per hydrogen molecule would be about
$\zeta_0 \approx 5 \times 10^{-17}$ sec$^{-1}$.  This value is
comparable to, but somewhat larger than, the ionization rates due to
cosmic rays in the interstellar medium (where typical estimates imply 
$\zeta\approx1-3\times10^{-17}$ sec$^{-1}$; e.g., \citealt{vandertak}).  
Since the star formation efficiency $\sfe$ is low, the amount of
material that the ejecta mix with is expected to be larger by a factor
of $1/\sfe\sim100$; on the scale of a star-forming region, the
ionization rate is thus smaller by this factor, so that we expect
$\zeta\sim10^{-19}$ sec$^{-1}$ (see Section \ref{sec:global}). Notice
also that one expects a wide range of values for the ionization rates
due to both SLRs and cosmic rays \citep{fam,cleevestts,cleevesslr}.

For sufficiently short spans of time, the contribution of SLRs to
ionization rates dominates over that of long-lived radioactive
species. For the abundance patterns deduced for the Solar Nebula, for
example, the contribution from long-lived nuclei is smaller by a
factor of $\sim10^4$ \citep{umenakano} at the start of the epoch
($t=0$). For the benchmark case illustrated by Figure
\ref{fig:ionize}, the short-lived nuclear species continue to dominate
until time $t\approx16$ Myr. As a result, we ignore the contribution
of long-lived nuclei for the remainder of this paper. Nonetheless,
this issue should be examined in the future. 

\section{Distributions of Nuclear Yields for Clusters} 
\label{sec:clustyield} 

\subsection{Yields for Clusters in the Large-N Limit} 
\label{sec:ystats} 

In the limit of large clusters, $N \to \infty$, the distributions of
nuclear yields will approach a gaussian form (see below). For this
regime, this section determines the mean values for the distributions
and their expected widths, where these quantities are a function of
stellar membership size $N$.

Here, the radioactive yield for a cluster with size $N$ 
is given by the sum 
\be
\yieldn = \sum_{j=1}^N \yieldj \, , 
\label{radiosum} 
\ee
where $\yieldj$ is the radioactive yield of species $A_k$ from the
$jth$ cluster member (with mass $m_j$). Only the massive stars (with
$m_j>8$) explode as supernovae at the end of their lives and
contribute to the radioactive yield of the cluster. As a result, the
quantity $\yieldj=0$ for most cluster members.

In this treatment, we assume that the radioactive yield for a given
star is determined by the stellar mass, which is drawn independently
from a specified stellar IMF. The sum from equation (\ref{radiosum})
is thus the sum of random variables. Here the variables are the
radioactive yields of the individual stars, so that the variables are
drawn from a known distribution, which is in turn determined by the
IMF and by stellar nucleosynthesis.  In the limit $N\gg1$, the
expectation value $\exyieldn$ of the radioactive yield for
the cluster is given by
\be 
\exyieldn = N \exstar\,,
\label{expectcn} 
\ee 
where $\exstar$ is the expectation value of the yield of radioactive
species $A_k$ per star, as defined via equation (\ref{radiomean}).  Keep
in mind that the radioactive yield for a cluster will converge to the
value implied by this expectation value in equation (\ref{expectcn})
only in the limit of large $N$.  The minimum value of cluster
membership $N$ required for this convergence is discussed below.
Small clusters often display large departures from the expectation
value.

In the limit of large $N\gg1$, the central limit theorem implies that
the distribution of yields $\yieldn$ must approach a gaussian form
(e.g., \citealt{feller}). In practice, however, the convergence is rather
slow. In addition, since only a small fraction of stars contribute to
the nuclear abundances, large $N$ values are necessary for
convergence.  One of the issues of interest here is the value of
stellar membership $N$ required for statistical considerations to be
valid. In the large $N$ limit, where the central limit theorem
applies, the resulting gaussian form for the composite distribution is
independent of the form of the initial distributions, i.e., it is
independent of the stellar IMF and the mass-luminosity relation. The
width of the distribution also converges to the value given by 
\be 
\sigbar^2 = {1 \over N}\sum_{j=1}^N \sigma_j^2 
\quad \Rightarrow \quad \sigbar = \sqrt{N} \sigma_\ast \,, 
\label{nvariance}
\ee 
where $\sigma_\ast$ is the width of the individual distribution
and is defined by 
\be 
\sigma_\ast^2 \equiv \exstarsqu - \exstar^2 \,.
\label{starvariance} 
\ee 

The expectation values $\exstar$ and widths $\sigma_\ast$ of the
distributions of radioactive yields are listed in Table 1 for the five
species of radionuclides considered in this paper (and for our chosen
form of the stellar IMF). Results for given for all five isotopes
using the WW models of stellar evolution, whereas results are only
given for $^{26}$Al and $^{60}$Fe using the LC models (results are not
available for the other nuclear species).  Both the expectation values
and the widths are given in units of $\mu M_\odot$.  For each
radioactive species, the expectation value $\exstar$ and the width
$\sigma_\ast$ of the distributions are given for three values of the
index $\gamma$ of the stellar IMF. The results are not overly
sensitive to the slope of the IMF in the mass range $m>8$, primarily
because the radioactive yields are not sensitive functions of
progenitor mass (see Figures \ref{fig:yieldalfe} and
\ref{fig:yieldclca}). However, the yields are directly proportional to
the fraction $\fsuper$ of stars above the supenova mass threshold; for
the cases shown in Table 1 we have used $\fsuper=0.005$.

The fifth column in Table 1 lists the number of stars $N_{X}$ in a
cluster required for the width of the distribution for the cluster to
be smaller than the expectation value of the yield of the cluster (for
the WW yields).  Since the expectation value of the yield for a
cluster is proportional to $N$ (see equation [\ref{expectcn}]) and the
width of the distribution is proportional to $\sqrt{N}$ (see equation
[\ref{nvariance}]), the benchmark cluster size
$N_{X}=(\sigma_\ast/\exstar)^2$. Notice also that this cluster size
$N_X$ is that necessary to make the distribution of yields narrower
than its expectation value. Even larger membership sizes $N$ are
required for the distribution to approach a pure gaussian form. 

Table 1 shows an interesting discrepancy between the results obtained
from the two nuclear models. The expectation values for the nuclear
yields per star show relatively moderate differences between the WW
and LC models (compare columns 3 and 6 for $^{26}$Al and $^{60}$Fe),
as expected from the results shown in Figures \ref{fig:yvgamma} --
\ref{fig:massratio}. However, the widths of the distributions (compare
columns 4 and 7) are much wider for the LC models than for the WW
models, especially for $^{60}$Fe.  Part of this difference arises
because the LC models provide results for larger progenitor masses,
and the yields increase with mass. The WW models end at $m=40$, and we
use the yields for the $m=40$ model for all higher masses; an
alternate extrapolation scheme could resolve part of this difference.
However, the LC models also show a steeper dependence of nuclear
yields with progenitor mass, especially for $^{60}$Fe (see Figure
\ref{fig:yieldalfe}). As shown in the following subsection, the wider
distributions for the LC models require larger clusters (in stellar
membership size $N$) to approach gaussian forms for the distribution
of radioactive yields per cluster.

\begin{table}
\begin{center}
\centerline{\bf Table 1: Parameters for Radio Isotope Distributions} 
\medskip 
\begin{tabular}{ccccccc} 
\hline 
\hline 
\hline 
Nuclear Species & $\gamma$ & ${\exstar}_{(WW)}$ & $\sigma_{\ast(WW)}$ 
& $N_{X}$ & ${\exstar}_{(LC)}$ & $\sigma_{\ast(LC)}$\\  
\hline 
\hline 
$\,$      & 1.3 &   0.212 &  4.18 &  388 & 0.333 & 7.82 \\
$^{26}$Al & 1.5 &   0.195 &  3.88 &  394 & 0.297 & 6.92 \\
$\,$      & 1.7 &   0.180 &  3.59 &  398 & 0.267 & 6.13 \\
\hline 
$\,$      & 1.3 &  0.0142 & 0.623 & 1940 \\
$^{36}$Cl & 1.5 &  0.0137 & 0.611 & 1990 \\
$\,$      & 1.7 &  0.0131 & 0.594 & 2060 \\
\hline
$\,$      & 1.3 &  0.0699 & 2.76 & 1560 \\
$^{41}$Ca & 1.5 &  0.0697 & 2.74 & 1540 \\
$\,$      & 1.7 &  0.0686 & 2.70 & 1550 \\
\hline
$\,$      & 1.3 &   0.329 &  7.75 &  556 \\
$^{53}$Mn & 1.5 &   0.326 &  7.63 &  549 \\
$\,$      & 1.7 &   0.318 &  7.44 &  549 \\
\hline
$\,$      & 1.3 &   0.179 &  3.52 &  387 & 0.213 & 10.1\\
$^{60}$Fe & 1.5 &   0.175 &  3.48 &  396 & 0.168 & 8.53\\
$\,$      & 1.7 &   0.169 &  3.42 &  409 & 0.132 & 7.16\\ 
\hline 
\hline 
\hline 
\end{tabular} 
\caption{The first column of the table gives the species of   
radionuclide and the second column lists the index of the stellar IMF.
In the third column, $\exstar$ is the expectation value of the
radioactive yield per star using the WW models, whereas $\sigma_\ast$
(fourth column) is the width of the corresponding distribution of
yields; both quantities are given in units of $\mu M_\odot$ = 
$10^{-6}M_\odot$. In the next column, $N_X$ is the number of stars in
a cluster required for the width of the distribution of yields for the
cluster to be smaller than the expectation value. For $^{26}$Al and
$^{60}$Fe, the table also lists the expectation value of the yield per
star and the corresponding width of the distribution for the LC
models. } 
\end{center} 
\end{table} 

\bigskip
\begin{table}
\begin{center} 
\centerline{\bf Table 2: Radio Isotope Properties for the Early Solar Nebula} 
\medskip
\begin{tabular}{ccccccc} 
\hline
\hline 
Nuclide & Daughter & Reference & Half-life & Mass Fraction & Mass & Uncertainty\\ 
$A_k$ & $D_k$ & $R_k$ & $\tau_{1/2}$ (Myr) & $X_k$ & $M_k$ $(pM_\odot)$ 
& $(\Delta M_k)/M_k$\\ 
\hline
\hline
$^{26}$Al& $^{26}$Mg& $^{27}$Al& 0.72& $3.8\times10^{-9}$& 190& 0.11\\  
$^{36}$Cl& $^{36}$Ar& $^{35}$Cl& 0.30& $8.8\times10^{-10}$& 44& 0.46\\  
$^{41}$Ca& $^{41}$K& $^{40}$Ca& 0.10& $1.1\times10^{-12}$& 0.055& 0.094\\  
$^{53}$Mn& $^{53}$Cr& $^{55}$Mn& 3.7& $4.0\times10^{-10}$& 20& 0.13\\  
$^{60}$Fe& $^{60}$Ni& $^{56}$Fe& 2.6& $1.1\times10^{-9}$ & 55& 0.35\\  
\hline  
\hline 
\end{tabular} 
\end{center}
\end{table} 

\begin{figure} 
\figurenum{9} 
{\centerline{\epsscale{0.90} \plotone{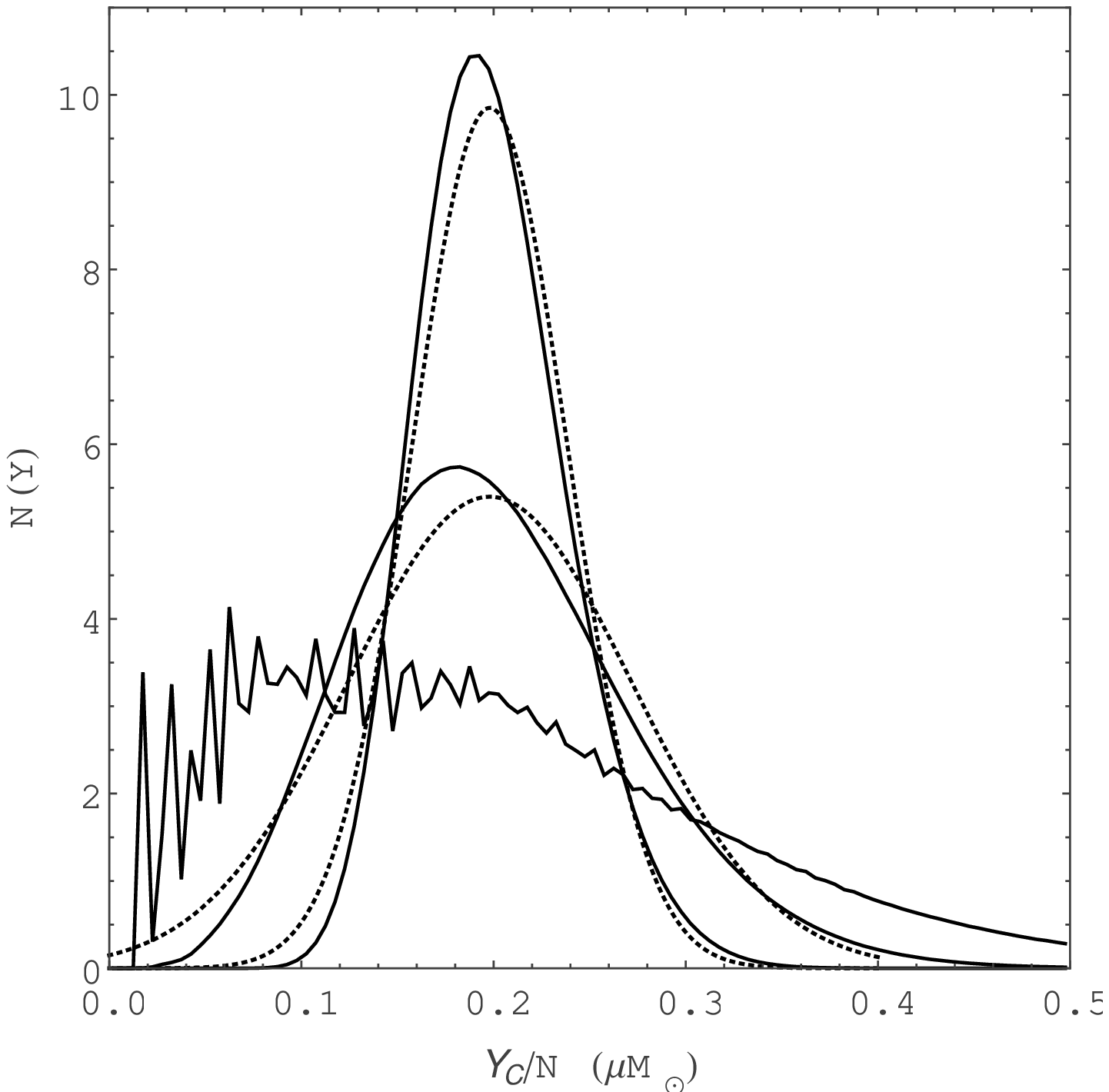} } } 
\figcaption{Distribution of radioactive yields of $^{26}$Al (using 
WW results) for clusters and $N=10^4$ (narrow curves), $N=3000$ (wider
curves), and $N=1000$ (widest, irregular curve). Solid curves show the
distributions obtained from sampling the IMF for a large collection of
clusters with fixed membership size $N$. Dotted curves show the
corresponding gaussian profile predicted analytically. The yields for
all distributions are scaled by the stellar membership size $N$. }
\label{fig:ldistal} 
\end{figure}  

\begin{figure} 
\figurenum{10} 
{\centerline{\epsscale{0.90} \plotone{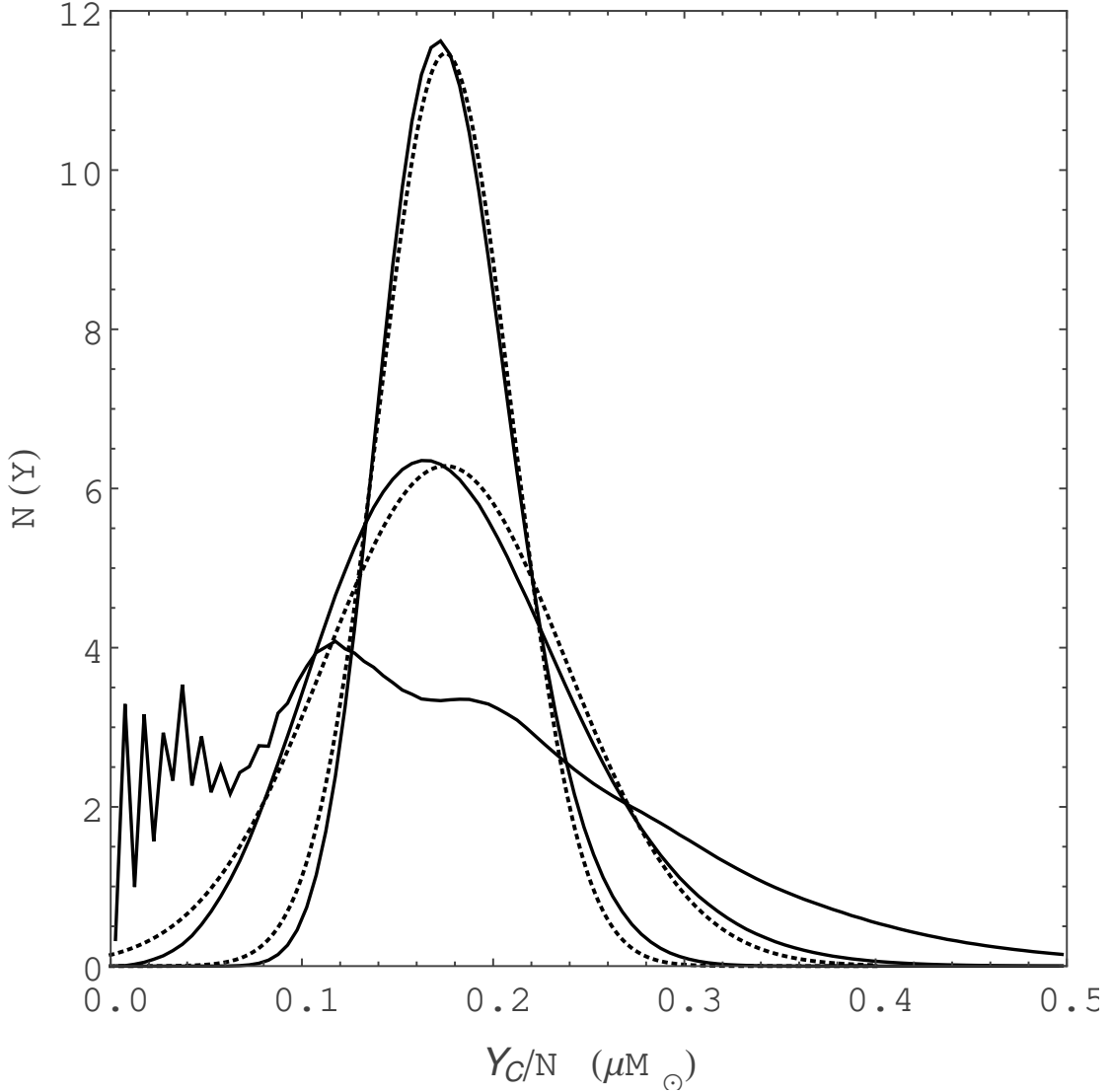} } } 
\figcaption{Distribution of radioactive yields of $^{60}$Fe (using 
WW results) for clusters and $N=10^4$ (narrow curves), $N=3000$ (wider
curves), and $N=1000$ (widest, irregular curve). Solid curves show the
distributions obtained from sampling the IMF for a large collection of
clusters with fixed membership size $N$. Dotted curves show the
corresponding gaussian profile predicted analytically. The yields for
all distributions are scaled by the stellar membership size $N$. }
\label{fig:ldistfe} 
\end{figure}  

\subsection{Yields for Clusters Determined via Sampling} 

Next we determine the distributions of nuclear yields for clusters by
direct numerical sampling of the stellar IMF. Figure \ref{fig:ldistal}
shows the distribution of radioactive yields of $^{26}$Al for clusters
with fixed stellar membership size $N=1000$, 3000, and $10^4$. The
solid curves show the distributions obtained by sampling the IMF for a
large number of clusters with fixed $N$, where the nuclear yields are
determined by the WW stellar models. Note that the radioactive yields,
shown on the horizontal axis, are scaled by the cluster size $N$ (so
that the peak and mean values are nearly independent of $N$).  For
comparison, the dotted curves show the gaussian profiles calculated 
using the mean value from equation (\ref{expectcn}) and the variance
from equation (\ref{nvariance}).  The relative widths of the
distributions decrease with increasing $N$, as expected (see Section
\ref{sec:ystats}). For the larger stellar membership sizes ($N=3000$
and $10^4$), the clusters contain enough massive stars so that the
distributions are close to the gaussian benchmarks. Nonetheless, the
true (sampled) distributions are slightly asymmetric, with the peak
value somewhat smaller than the expectation value. Although these
departures are small for large $N$, clusters with smaller membership
$N$ display large departures from gaussian profiles. The figure also
shows the result obtained by sampling clusters with only $N=1000$
members (shown as the irregular, wide curve). In this case, many
clusters only have one or two massive stars large enough to explode,
so that the low end of the distribution shows a great deal of
structure (which reflects the irregular structure of the radioactive
yields as a function of progenitor mass, as shown in Figure
\ref{fig:yieldalfe}). Although the spikey nature of the distibution
(for $N=1000$) is visually prominent, perhaps the most important
departure from from a gaussian form is the asymmetry toward lower
values. As expected, in the opposite limit where $N\to\infty$, the
distributions approach true gaussian forms.

Figure \ref{fig:ldistfe} shows the corresponding distributions of
yields for $^{60}$Fe, again using the WW nuclear models and for
cluster membership sizes $N$ = 1000, 3000, and $10^4$. These
distributions are analogous to those obtained for $^{26}$Al (compare
with Figure \ref{fig:ldistal}).  As expected, the larger clusters
(with $N=3000$ and $10^4$) display nearly gaussian profiles, as shown
by the dotted curves in the figure. On the other hand, slightly
smaller clusters with $N$ = 1000 show complicated, irregular
structure, for the same reasons discussed above. The distributions are
also asymmetric, with the peak value smaller than the mean value; this
trend is small for $N \ge 3000$, but significant for $N=1000$. 

\begin{figure} 
\figurenum{11} 
{\centerline{\epsscale{0.90} \plotone{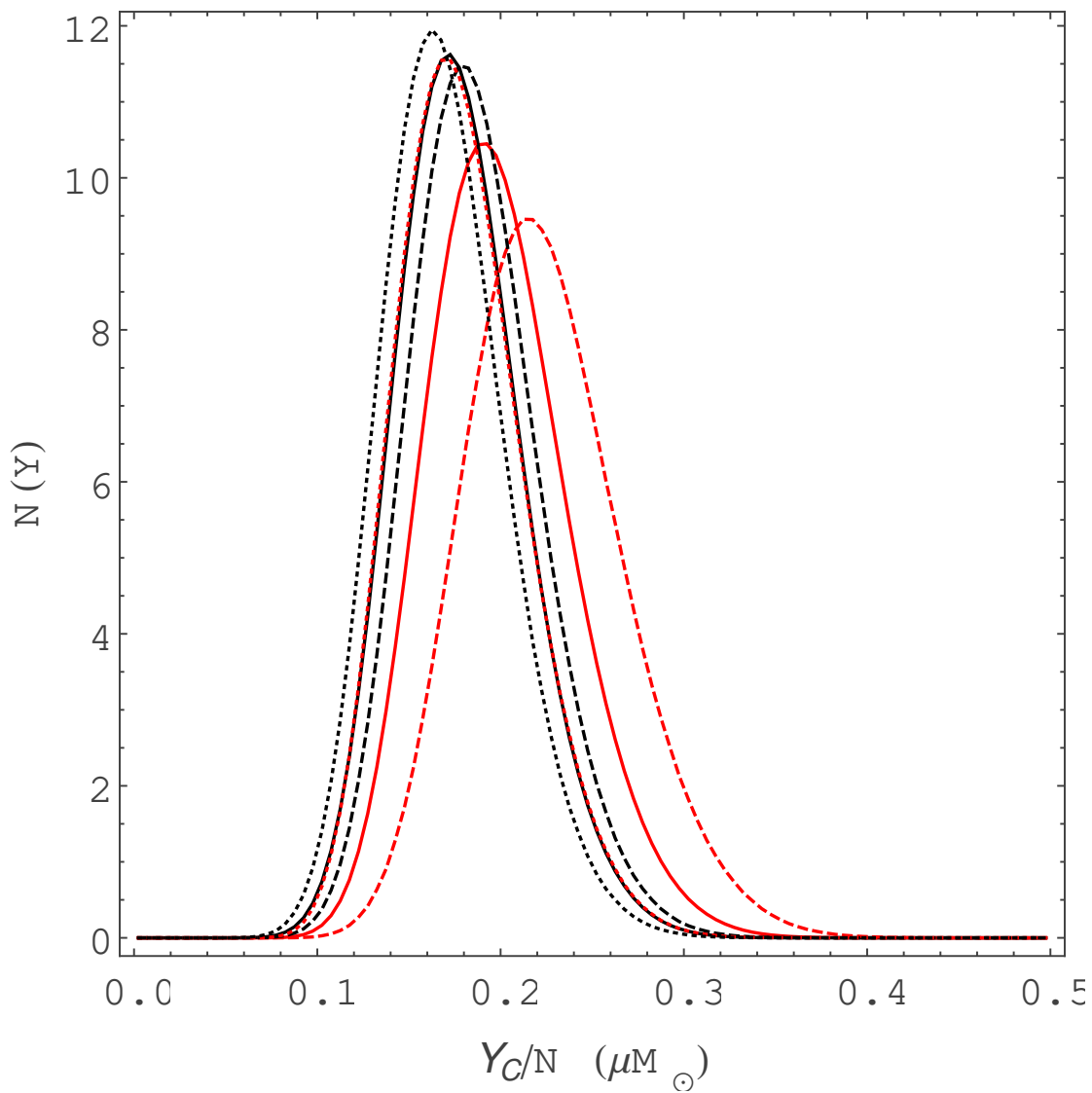} } } 
\figcaption{Distribution of radioactive yields for difference values  
of the index $\gamma$ of the stellar IMF. Distributions are shown for
both $^{26}$Al (red curves) and $^{60}$Fe (black curves) using the WW
nuclear model. The cluster size is taken to $N=10^4$. For each
isotope, distributions are shown for three values of the index
$\gamma$ = 1.75 (dotted curves), 1.5 (solid curves), and 1.25 (dashed
curves). }
\label{fig:distvgam} 
\end{figure}

Next we consider the effect of the stellar IMF on the resulting
distributions of nuclear yields.  Figure \ref{fig:distvgam} shows
the distributions of yields for clusters with $N=10^4$ stars and for
the WW nuclear models. Results are shown here for both $^{26}$Al (red
curves) and $^{60}$Fe (black curves). For each isotope, distributions
for shown for three choices of the index $\gamma$ of the stellar IMF,
where $\gamma$ = 1.25 (dashed curves), 1.5 (solid curves), and 1.75
(dotted curves). The effect of varying the index $\gamma$ is modest:
The distributions shift their mean values slightly as $\gamma$ varies,
as expected given the dependence of the expectation values shown in
Figure \ref{fig:yvgamma}. The expected yields per star decrease with
increasingly index $\gamma$, so that the distributions move to the
left as the stellar IMF becomes steeper. For clusters with different
stellar membership sizes $N$ (not shown), one obtains analogous
distributions; they are relatively wider for smaller $N$ and narrower
for larger $N$, as illustrated in Figures \ref{fig:ldistal} and
\ref{fig:ldistfe}. 

\begin{figure} 
\figurenum{12} 
{\centerline{\epsscale{0.90} \plotone{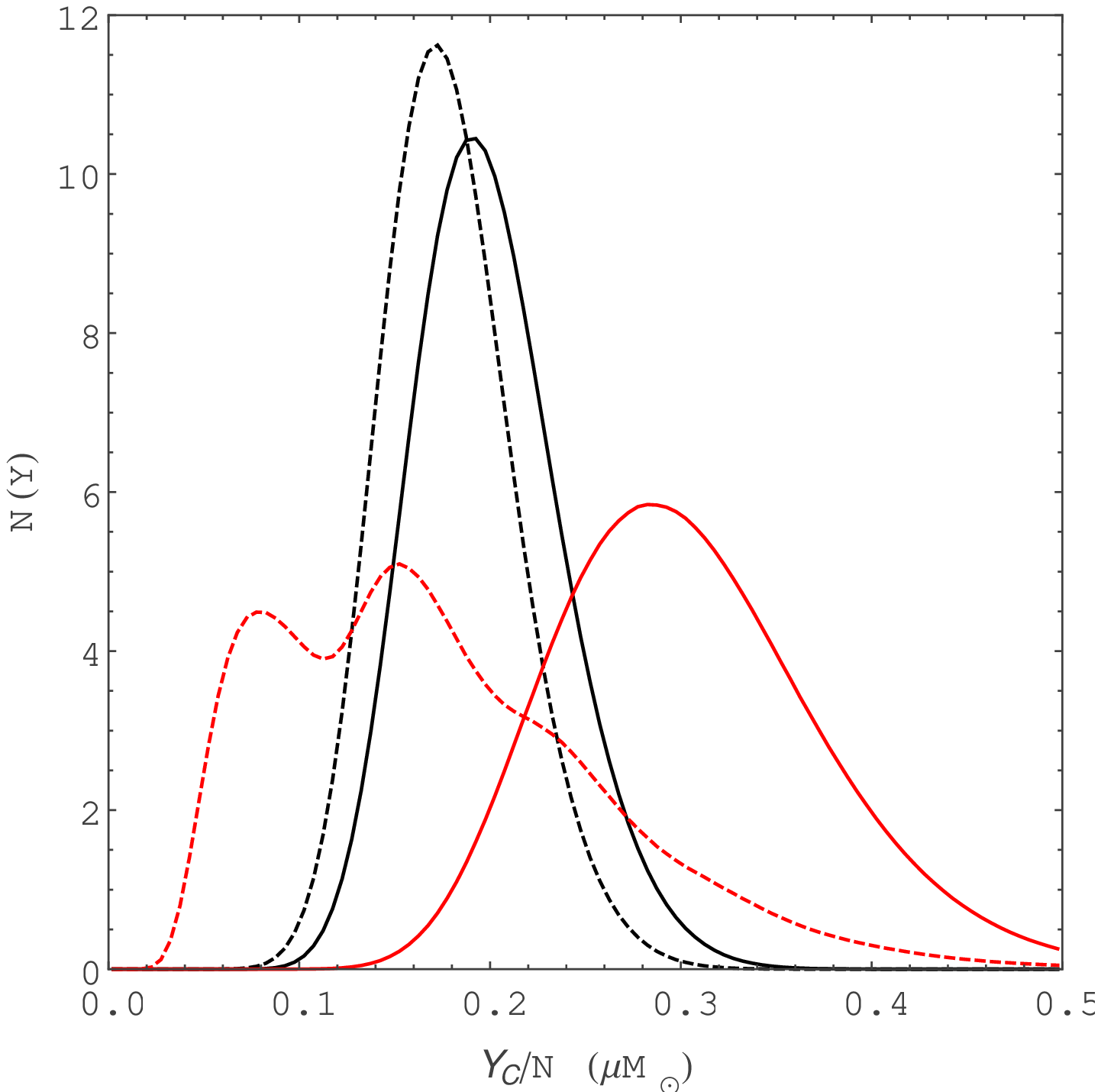} } } 
\figcaption{Comparison of the two nucleosynthesis models for the  
distribution of radioactive yields in clusters with $N=10^4$.  The 
black curves show the distributions for the WW models, whereas the red
curves show the distributions for the LC models. The distributions of
yields for $^{26}$Al are shown as the solid curves, whereas the
distributions for $^{60}$Fe are shown as dashed curves. }  
\label{fig:distvmodel} 
\end{figure}   

Figure \ref{fig:distvmodel} compares the two models for stellar
nucleosynthesis used in this work. Here we present distributions of
nuclear yields for $^{26}$Al and $^{60}$Fe, as determined using both
the WW and LC results. The cluster membership size is taken to be
$N=10^4$ for this comparison. For the WW models, the distributions are
relatively narrow and approach gaussian forms, as depicted by the
black curves in the figure (and as shown previously).  For the LC
models, however, the distributions are markedly wider, as depicted by
the red curves. This behavior is in keeping with the larger widths
$\sigma_\ast$ given in Table 1.  For $^{26}$Al, the distribution for
the LC yields is shifted to the right compared to that for the WW
yields --- consistent with the larger expectation value found using
the LC models --- and is close to gaussian. For $^{60}$Fe, however,
the distribution retains a significantly non-gaussian form, even for
this relatively large stellar membership size $N$. For the LC models,
we thus find that larger clusters (larger $N$) are required for the
distributions of nuclear yields to become gaussian, with the effect
more pronounced for $^{60}$Fe. This behavior results from the steep
dependence of the $^{60}$Fe yields with progenitor mass (see Figure
\ref{fig:yieldalfe}) coupled with the steepness of the stellar IMF: A
few, rare large stars can contribute an enormous amount of $^{60}$Fe,
so that large stellar populations (large $N$) are required to fully
sample the distribution.  In this case, the stellar membership size
required for a gaussian distribution is larger than $N=10^4$.

\subsection{Expectation Values} 
\label{sec:exvalues} 

In this section we find the expectation values for various quantities
of interest.  We first consider the expectation value for the
radioactive yield $Y_C$ of an entire cluster. To obtain this quantity,
we integrate over the distribution of cluster sizes and the
distribution of possible radioactive yields (per cluster): 
\be
\langle Y_C \rangle = 
\int_1^{N_{max}} {C_N \over N^2} dN \int_0^\infty dY = 
\int_1^{\nmax} {C_N \over N^2} dN N \exstar = C_N \exstar \log \nmax \, . 
\ee
Since $C_N\approx1$ and $\nmax\approx10^6$, we find that 
\be
\langle Y_C \rangle \approx 6\log10 \exstar \,, 
\ee
where $\exstar$ is the radioactive yield per star for a given 
nuclear species (as given in Table 1). For example, a ``typical'' 
cluster produces only about $2.8 \times 10^{-6} M_\odot$ of 
$^{26}$Al. This value is small because the typical cluster is 
small: The expectation value for the cluster membership size 
is given by 
\be
\langle N \rangle = 
\int_1^{N_{max}} {C_N \over N^2} N dN = C_N \log \nmax \approx 14.
\ee
This expectation value is small because we take the distribution of
cluster membership sizes to extend all the way down to $N=1$.  The
resulting distribution will thus have many small clusters, which leads
to the small value of $\langle{N}\rangle$. These small clusters
contain only a small fraction of the stellar population, however, so
that most stars reside in much larger clusters.

A related quantity is the radioactive yield per cluster that a typical
star experiences within its birth cluster. This quantity, denoted here
as $\langle Y_{C\ast} \rangle$, is given by 
\be
\langle Y_{C\ast} \rangle = 
\int_1^{N_{max}} {C_P \over N} dN \int_0^\infty dY = 
\int_1^{\nmax} {C_P \over N} dN N \exstar = C_P \exstar \nmax \, . 
\ee
In this case, $C_P = 1/\log\nmax$ and we obtain the estimate
\be
\langle Y_{C\ast} \rangle = {\nmax \over \log \nmax} \exstar 
\approx 7.2 \times 10^4 \exstar \approx 0.0145 M_\odot \,.
\label{typicalstar} 
\ee

As considered in the next section, only a small fraction of the
nuclear yield from a cluster will be delivered to any given solar
system.  Before considering that issue in detail, however, it is
useful to obtain a rough estimate: If we use a typical distance of a
solar system to the cluster center of $d\sim1$ pc, the expected
fraction of the nuclear material that is intercepted by a disk (with
radius 30 AU) is only about $f\sim3\times10^{-9}$.  If a typical star
is born in a cluster with nuclear yield described by equation
(\ref{typicalstar}), the expected mass of $^{26}$Al impinging on a
typical star/disk systems is thus about $4\times10^{-11}M_\odot$.  
For comparison, the estimated mass fraction of $^{26}$Al in the early
Solar Nebula is $X=4 \times 10^{-9}$, so that the mass of $^{26}$Al is
about $2 \times 10^{-10} M_\odot$ (five times larger than the
canonical value --- see below). Although this argument uses only typical values, 
it suggests that clusters can provide nuclear enrichment to their
constituent solar systems at levels comparable to (but often somewhat
less than) those estimated for our Solar Nebula.

\section{Distributions of SLR Yields Delivered to Solar Systems} 
\label{sec:systemyield}  

A typical star in a typical cluster will intercept a only fraction $f$
of the radioactive yield produced by the entire cluster. To start, we
ignore timing issues. In this limiting case, the fraction $f$ is given
by the geometrical factor 
\be
f = f(r) = {\pi r_d^2 \over 4 \pi r^2} \cos\theta \,,
\ee
where $r_d$ is the disk radius, and $r$ is the distance from the solar
system to the cluster center (where the high mass stars, and hence the
supernova ejecta, originate). The factor of $\cos\theta$ takes into
account the fact that the disk is not, in general, facing the
supernova blast wave; the distribution of angles is expected to be 
uniform in $\mu\equiv\cos\theta$ with a mean of 1/2. The radius $r$ 
must be larger than the radius for which the disk (with disk radius 
$r_d$) is stripped due to the blast; for a disk radius $r_d$ = 30 AU,
and for typical supernova energies, this minimum radial distance 
$r_{min}\approx0.1$ pc \citep{chevalier,ouellette,adams2010}.

In this treatment of the problem, we assume that the supernova ejecta
are distributed uniformly and isotropically, and are not impeded
before reaching the target protoplanetary disks. This scenario is thus
idealized, as supernova ejecta can be clumpy \citep{grefenstette} and
intervening molecular cloud material can interfere with nuclear
delivery (e.g., see \citealt{sashida}).

A great deal of previous work has considered the problem of injecting
SLRs from supernova ejecta into existing circumstellar disks.  
The efficiency of injection ranges from essentially 100 percent  
for injection into circumstellar disks \citep{looney,ouellette}, down
to only about 10 percent for injection into pre-collapse clouds
\citep{vanhala}. For injection into disks, such high efficiencies require
the SLRs to condense onto dust grains before striking the disk; more
specifically, recent work finds that the efficiencies can be 70
percent or higher only if the grain sizes are larger than $\sim0.4\mu$m 
(see \citealt{ouellette2010}, in particular their Figure 6).
Unfortunately, however, the expected size of grains produced during
supernovae remain uncertain. A recent study advocates dust grains with
sizes smaller than $\sim0.1\mu$m \citep{bianchi}, which would lead to
lower injection efficiencies. In this work, we assume 100 percent
efficiency, but the results can be scaled (downward) for any choice of
this parameter. Another important issue is the transport of the SLRs
after they are acquired; simulations carried out to date suggest that
mixing is indeed efficient \citep{boss2011,boss2013}.

\subsection{Expectation Values for SLR Delivery} 

The expectation value $\langle{Y_\ast}\rangle$ for the yield of SLRs
intercepted by a given solar system takes the form 
\be
\langle Y_{\ast} \rangle = 
\int_1^{N_{max}} {C_P \over N} dN \int_0^\infty dY 
\int_{r_{min}}^R {dP \over dr} f(r) dr \,,
\ee
which can be written 
\be
\langle Y_{\ast} \rangle = 
\int_1^{\nmax} {C_P \over N} dN N \exstar 
\int_{r_{min}}^R {3-p \over R} \left( {r \over R} \right)^{2-p}
{\pi r_d^2 \over 8 \pi r^2} dr \,,
\ee
where we have used equation (\ref{dpdr}) to specifiy the probability
distribution of the radial positions and we have taken the mean value
of $\cos\theta$. This result can be simplified to the form 
\be
\langle Y_{\ast} \rangle = C_P \exstar 
{(3-p) r_d^2 \over 8} \int_1^{\nmax} {dN \over R^{3-p} } 
\int_{r_{min}}^R {dr \over r^p} \,. 
\ee
We can evalute the above result for any value of $p$ and any form for
the cluster radius function $R(N)$, which is specified by power-law
index $\alpha$ (from equation [\ref{rvsn}]). For the canonical choices 
$p$ = 3/2 and $\alpha$ = 1/3, we obtain
\be
\langle Y_{\ast} \rangle \approx C_P \exstar 
{3r_d^2 \over 4} {r_{min}}^{-1/2} R_0^{-3/2} N_0^{1/2} \nmax^{1/2} 
\approx 6.3 \times 10^{-5} \exstar \,,
\ee
where we have used typical values to obtain the final approximate
equality. Since the yields of both $^{26}$Al and $^{60}$Fe are of
order 0.2 $\mu M_\odot$ (see Figure \ref{fig:yvgamma} and Table 1), 
the typical yield for these SLRs is about 10 -- 20 $pM_\odot$ 
(or $1-2\times10^{-11}M_\odot$). 

The expectation values for SLR yields discussed here are comparable
to --- but somewhat smaller than --- the SLR masses that are inferred
for the early Solar Nebula. For comparison, Table 2 lists the isotopes
of interest for this paper, along with the daughter products, the
reference isotopes, the half-lives, the mass fractions, and the total
masses. The abundances listed in the table are inferred from
meteoritic data, which has been compiled by numerous previous authors
(e.g., see \citealt{umenakano,looney,young,dauphas}; and references
therein).  For each SLR, the total mass is estimated from the mass
fraction, where we assume a typical mass for the Solar Nebula of
$M_d=0.05M_\odot$. We note that the abundance for $^{60}$Fe listed
here (taken from \citealt{tachibana}) has been re-measured by other
workers \citep{tang}, who found lower abundances. In any case, the
total SLR masses listed in Table 2 are somewhat larger than the
expectation values discussed above. Taken at face value, this finding
implies that Solar System abundances result from the high end of the
distribution of possible values. To asssess the probabilties, we need
the full distribution, which is determined in the next subsection.

\subsection{Distributions for SLR Delivery} 

Using the probability distributions for cluster yields and for the
radial positions of stars within clusters, we can find the
distribution of radioactive yields delivered to the constituent solar
systems.  Let $Y_C$ be the nuclear yield for a given cluster, where
the value of $Y_C$ is distributed according to a probability
distribution $dP/dY$.  Let $\xi=r/R$ be the radial position of the
recipient solar system within the cluster, where $R$ is the cluster
radius; the position is distributed according to $dP/d\xi$, which
depends on the density profile of the cluster. The mass $M_{ss}$ of
radioactive material delivered to a solar system is given by 
$M_{ss}=Y_C f(r)$, which can be written as 
\be
M_{ss} = Y_C {\pi r_d^2 \over 4 \pi R^2 \xi^2 } \mu \,,
\label{mpsystem} 
\ee
where $\mu=\cos\theta$. To find the distribution of the mass $M_{ss}$
of intercepted nuclear material, we need to specify the distributions
$dP/dY$, $dP/d\xi$, and $dP/d\mu$. The projection factor $\mu$ has
uniform-random distribution ($dP/d\mu=1$) on the interval [0,1]. To
start, however, we ignore projection effects by setting $\mu=1$
(projection effects will be reinstated later).

As shown above, for clusters with sufficiently large stellar
membership size $N$, the distribution of yields is nearly gaussian,
i.e.,
\be
{dP \over dY} = \aconst \exp \left[ - 
{(Y_C - \langle Y \rangle)^2 \over 2 \sigma^2}\right] \,,
\ee
where $\aconst$ is the normalization constant and is given by 
\be
\aconst = {2 \over \sqrt{2\pi} \sigma } \left[ 1 + 
\erf \left( {\langle Y \rangle \over \sqrt{2} \, \sigma} \right) 
\right]^{-1} \,,
\label{aconst} 
\ee 
where $\erf(z)$ is the error function \citep{abrasteg}. Further, the
expectation value $\langle{Y}\rangle$ and the width $\sigma$ of the
distribution are given by 
\be
\langle Y \rangle = N \langle Y \rangle_\ast 
\qquad {\rm and} \qquad \sigma = \sqrt{N} \sigma_\ast \,,
\ee
where the quantities with starred subscripts denote the values
per star, calculated from convolving the nuclear yields with 
the stellar IMF. 

As discussed earlier, the distribution of radial positions depends
on the cluster density profile, i.e., 
\be
{dP \over d\xi} = (3-p) \xi^{2-p} \,,
\ee
where the index $p$ of the cluster density profile lies in the range
$1 \le p \le 2$. Note that the radial position has a uniform
(constant) distribution for the choice $p=2$.

Next we define a new variable 
\be 
y \equiv {Y_C \over \langle Y \rangle} \,,
\ee
so that the mass $M_{ss}$ can be written 
\be
M_{ss} = \langle Y \rangle {\pi r_d^2 \over 4 \pi R^2} 
{y \over \xi^2 } = M_0 X \,,
\label{mssx}
\ee
where the second equality defines the composite variable 
\be
X \equiv {y \over \xi^2} \,,
\ee
and the benchmark scale 
\be 
M_0 \equiv \langle Y \rangle {\pi r_d^2 \over 4 \pi R^2}\,.
\label{benchmass} 
\ee
For typical values we obtain 
\be
M_0 = 0.84 pM_\odot 
\left({N \over 3000} \right) 
\left({\langle Y \rangle_\ast \over 0.2 \mu M_\odot}\right)
\left({r_d \over 30\,{\rm AU}}\right)^2 
\left({R \over 2\,{\rm pc}}\right)^{-2}\,.
\ee
The benchmark mass scale is thus of order $1pM_\odot=10^{-12}M_\odot$;
further, cluster radii are observed to vary as $R \propto N^{\alpha}$,
where $1/4 \le \alpha \le 1/2$, so that this mass scale is a slowly
varying function of stellar membership size $N$. For comparison, the 
inferred abundances of SLRs for the early Solar Nebula correspond to 
masses in the range 20 -- 200 $pM_\odot$ (Table 2). Note that the 
expected value for the enrichment mass is given by the expectation 
value $M_0\langle{X}\rangle$, which will be larger than $M_0$. We 
can evalulate the quantity $\langle{X}\rangle$ to obtain
\be
\langle{X}\rangle = \int_0^\infty X {dP \over dX} dX = 
{(3-p) \over (p-1)} 
\left[ {\xi_{min}^{-(p-1)} - 1 \over 1 - \xi_{min}^{3-p}} \right] 
\left[ 1 + B \lambda^2 \exp (-1/2\lambda^2) \right] \,,
\label{xexpect1} 
\ee
where the constants $B$ and $\lambda$ are defined below. 
We also have introduced a minimum radius $\xi_{min}=r_{min}/R$ for
the solar system location within the cluster, and adjusted the
normalization constant accordingly.  Without this cutoff, the solar
system could lie arbitrarily close to the enrichment sources and the
integral would diverge. In practice, solar systems that are too close
to the cluster center, where the supernova explosions occur, will have
their disks destroyed; as a result, blast wave physics enforces a
minimum radius of $0.1-0.2$ pc \citep{chevalier,ouellette,adams2010},
which implies $\xi_{min}\sim0.05-0.20$. Under most circumstances,
$\lambda\ll1$ and $\xi_{min}\ll1$; in this limit, the expectation
value reduces to the simpler form 
\be
\langle{X}\rangle \to {(3-p) \over (p-1)} {1 \over \xi_{min}^{p-1}} \,. 
\label{xexpect2} 
\ee
This expectation value thus has a typical value $\langle{X}\rangle\sim10$.  
The corresponding expected value for radioactive mass enrichment is
thus about 10 $pM_\odot$, roughly comparable to, but still somewhat
smaller than, the levels inferred for the early Solar Nebula.

The distribution for the scaled variable $y$ takes the form 
\be
{dP \over dy} = \aconst \langle Y \rangle 
\exp \left[ - {\langle Y \rangle^2 \over 2 \sigma^2}
(y-1)^2 \right] \,,
\ee
where the normalization constant $\aconst$ is given by equation 
(\ref{aconst}). If we define the quantity 
\be
\lambda \equiv {\sigma \over \langle Y \rangle} = 
{\sigma_\ast \over \sqrt{N} \exstar} \,, 
\label{lambdadef} 
\ee
the distribution simplies to the form 
\be
{dP \over dy} = B 
\exp \left[ - {(y-1)^2 \over 2 \lambda^2}\right] \,,
\ee
where the normalization constant is given by 
\be
B = {2 \over \sqrt{2\pi} \lambda} \left[ 1 + 
\erf \left( {1\over \sqrt{2} \, \lambda} \right) \right]^{-1} \,.
\ee

Now we need to determine the cummulative probability $P(X)$ for the
variable $X$. The probability is given by the double integral 
\be
P(X) = \int_0^1 d\xi {dP \over d\xi} \int_0^{X \xi^2} 
{dP \over dy} dy \, ,
\ee
which can be written in the form 
\be
P(X) = \int_0^1 d\xi (3-p) \xi^{2-p} 
\int_0^{X \xi^2} dy \,B\, 
\exp \left[ - {(y-1)^2 \over 2 \lambda^2}\right] \,.
\ee
The differential probability is then given by 
\be
{dP \over dX} = (3-p) B \,
\int_0^1 d\xi\,\xi^{4-p} \exp \left[ - 
{({X \xi^2}-1)^2 \over 2 \lambda^2}\right] \,.
\label{dpdx} 
\ee
For large $X\gg1$, this result reduces to the power-law form 
\be
{dP \over dX} \approx {(3-p) \over 
1 + \erf \left[ \sqrt{2}/(2\lambda) \right] } X^{-(5-p)/2} 
{1 \over \lambda \sqrt{2\pi}} \int_0^\infty u^a du {\rm e}^{-(u-1)^2/2\lambda^2}
\equiv C X^{-(5-p)/2} \,,
\ee
where the second equality defines the normalization constant $C$
and where we have defined $a = (3-p)/2$. In the limit where 
$\lambda \ll 1$, the expression simplifies to the form 
\be
{dP \over dX} = C X^{-(5-p)/2} \qquad {\rm where} \qquad 
C = {1 \over 2} (3-p) \,. 
\label{dpdxlimit} 
\ee

One quantity of interest is the fraction of solar systems that will be
enriched in SLRs above a given threshold specified by $X_\ast$.  Note
that the mass of SLRs is given by $M = M_0 X$, where the benchmark
mass is of order $1pM_\odot$. The fraction of systems that receive
$X>X_\ast$ ($M>M_\ast$) can be written in the form 
\be
P(X>X_\ast) = {C \over (3-p)/2} X_\ast^{-(3-p)/2} \approx 
X_\ast^{-(3-p)/2} \,. 
\ee

The treatment thus far has neglected projection effects. Since
circumstellar disks are not, in general, aligned toward the supernova
ejecta, the distribution derived above must be convolved with the
distribution of orientation angles.  In practice, we have the
probability distribution $dP/dX$ for the variable $X$ that specifies 
the SLR masses delivered to individual solar systems (where $M=M_0X$).
We need to find the corresponding probability distribution $dP/dZ$ for
the composite variable $Z\equiv \mu X$, where $\mu=\cos\theta$ is
distributed uniformly on [0,1]. The cummulative probability is then
given by the integral 
\be
P(Z>Z_\ast) = \int_0^1 d\mu \int_{Z_\ast/\mu}^\infty {dP \over dX} dX \,.
\ee
Since we are interested in the regime where $X,Z \gg 1$, we can use
the limiting form given by equation (\ref{dpdxlimit}) to specify
$dP/dX$; as a result, the expression becomes
\be
P(Z>Z_\ast) = \int_0^1 d\mu \int_{Z_\ast/\mu}^\infty 
C X^{-(5-p)/2} dX \,, 
\ee
which can be evaluated to obtain 
\be
P(Z>Z_\ast) = {2 \over 5-p} Z_\ast^{-(3-p)/2}\,.
\label{pzcum} 
\ee
As a result, the fraction of solar systems that are enriched at a
given level is reduced by a factor of $2/(5-p)\approx0.5-0.67$ due to
projection effects. The corresponding differential probability for 
the variable $Z$ then becomes 
\be
{dP\over dZ} = {3-p \over 5-p} Z^{-(5-p)/2} \,. 
\ee

To fix ideas, we use the index $p=3/2$ and require that
$Z=\mu{X}>100$, which corresponds to nuclear enrichment at levels
comparable to those inferred for the early Solar Nebula. Using
equation (\ref{pzcum}), the fraction of solar systems that are
enriched at this level is found to be $P(Z>100)\approx0.018$.  For the
extreme values $p$ = 1 (2), we obtain $P(Z>100)$ = 0.005 (0.067). This
result suggests that only a few percent of all circumstellar disks are
enriched at levels comparable to our Solar System (similar results are
advocated by \citealt{parker}). However, for somewhat lower enrichment
levels (say, $Z>10$), the fraction increases to $\sim10$ percent for
clusters with $p=3/2$ and to $\sim21$ percent for clusters with $p=2$.
For comparison, after taking into account the projection factor, the
expectation value is about $\langle{Z}\rangle\sim5$ (see equations
[\ref{mssx} -- \ref{xexpect2}]). Note that this expectation value is
larger than the median because the distribution has a long tail
corresponding to high values of enrichment.

\begin{figure} 
\figurenum{13} 
{\centerline{\epsscale{0.90} \plotone{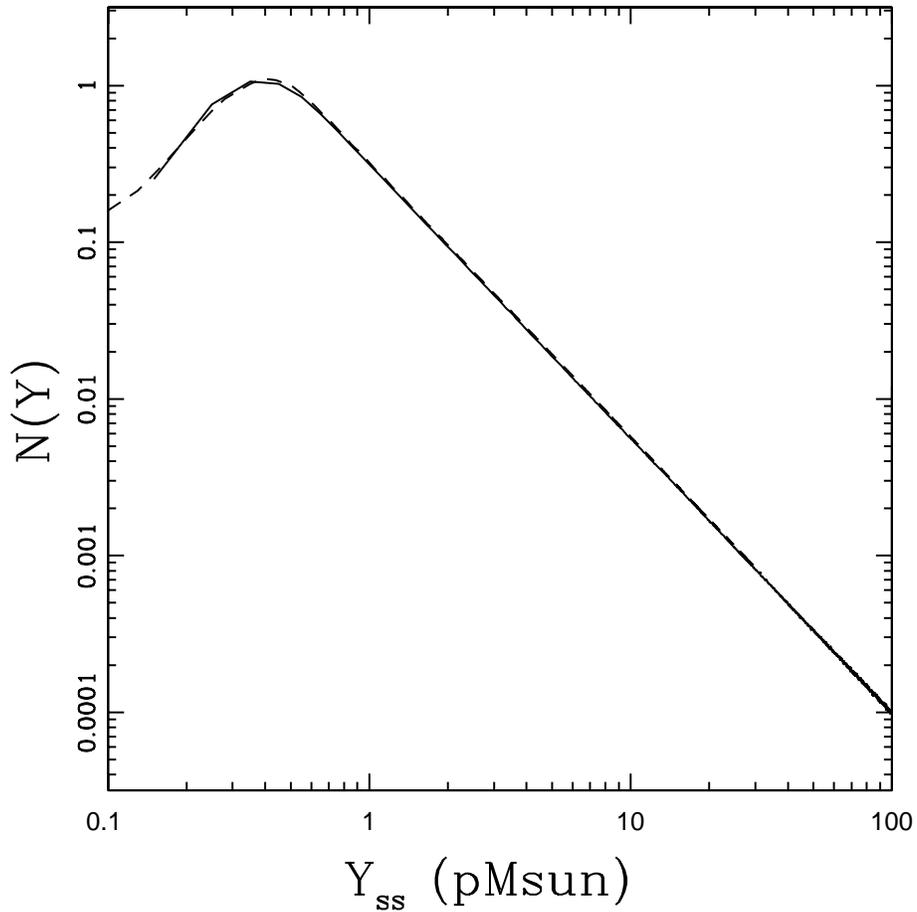} } } 
\figcaption{Distribution of total mass in $^{26}$Al delivered to solar   
systems living in clusters with stellar membership size $N$ = 3000.
Solid curve shows the composite distribution calculated by numerical
sampling both the stellar IMF and the radial positions within the
cluster (using density profile with index $p$ = 3/2). The dashed curve
shows the analytic distribution given by equation (\ref{dpdx}) with
benchmark mass scale given by equation (\ref{benchmass}). }
\label{fig:ss3000} 
\end{figure} 

This semi-analytic treatment works well for sufficiently large
clusters. For example, Figure \ref{fig:ss3000} shows the distribution
of the total mass in $^{26}$Al that is delivered to constituent solar
systems in clusters with membership size $N$ = 3000. The solid curve
shows the distribution obtained from numerical sampling, whereas the
dashed curve shows the result obtained from equations (\ref{dpdx}) and
(\ref{benchmass}). To obtain the numerical curve, we construct a
theoretical sample of one million clusters with $N$ = 3000. For each
cluster, the stellar IMF is sampled $N=3000$ times; for the massive
stars, the yield of $^{26}$Al is then determined using the WW models.
With the total supply of $^{26}$Al specified for the cluster, the
amount delivered to individual solar systems is determined from
equation (\ref{mpsystem}) after sampling the distribution of radial
positions, as given by equation (\ref{dpdr}). For both distributions,
the cluster radius is given by equation (\ref{rvsn}) and the index of
the cluster density distribution is $p$ = 3/2. The semi-analytical
result is in excellent agreement with the numerically determined
distribution, provided that the same parameters are used. 

The distributions shown in Figure \ref{fig:ss3000} are subject to
additional uncertainties. For example, we have used the cluster
radius from equation (\ref{rvsn}), which implies $R\approx3.16$ pc.
Some clusters could have slightly smaller radii (say, $R\sim2$ pc),
which would increase the SLR masses by a factor of $\sim$5/2. In
addition, the yields of SLR from the stellar evolution models are
uncertain (see Figures \ref{fig:yieldalfe}, \ref{fig:yvgamma},
\ref{fig:yvmass}, and \ref{fig:yvtime}).  Fortunately, however, the
semi-analytic treatment allows us to scale the whole distribution for
any choice of supernova yields --- one just needs to scale the
benchmark mass scale $M_0$ given by equation (\ref{benchmass}) for the
preferred choice of expectation value, as well as specify the width of
the distribution $\lambda$ using equation (\ref{lambdadef}).
Similarly, we can find the distributions for other SLR species (e.g.,
$^{60}$Fe) by using the appropriate values of $M_0$ and $\lambda$.
Finally we note that the SLR abundances quoted here do not  
include radioactive decay, so they represent the starting abundances,
immediately after injection. The subsequent decay results in a
reduction factor, e.g., as shown in Figure \ref{fig:ionize} as a
function of time since the supernova explosion (see also Section
\ref{sec:time} and the discussion of Section \ref{sec:conclude}). 

\section{Global Considerations} 
\label{sec:global} 

The radioactive yields considered here produce observable consquences
on larger scales.  We note that production of SLRs has been considered
previously on galactic scales (e.g., \citealt{timmesa,timmesb,diehl})
and in molecular clouds (e.g., \citealt{gounelle2008}). This section
briefly revisits the issue to see how observations on these larger
scales can constrain the distributions of SLR enrichment considered in
previous sections.

Let $\Gamma_{SF}$ be the star formation rate of a well-defined system.
Here we consider the ``system'' to be either the entire Galaxy or, on
a smaller scale, a molecular cloud complex.  For a given system, the
time evolution of the supply of radioactive elements, of species
$A_k$, is given by 
\be
{d M[A_k] \over dt} = {1 \over \mbar} \Gamma_{SF} \exstar - 
{\ln 2 \over \halflife} M[A_k] \,,
\label{mdotsupply} 
\ee
where $\halflife$ is the half-life; note that the star formation rate
is given in units of mass per unit time (rather than the number of
stars per unit time).  The steady-state condition, which sets the
equilibrium abundance, can then be written in the form 
\be
M[A_k] = {\halflife \Gamma_{SF} \exstar \over (\ln 2) \mbar} \,.
\label{mequilib} 
\ee 
This simple treatment assumes that the radioactive nuclei decay, but
cannot leave the system by other channels. Since the half-lives of
interest are short, of order 1 Myr, this assumption should be valid
for the Galaxy as a whole. For molecular clouds, however, supernovae
may not inject all of their SLRs into the cloud, so that losses can
occur.

\subsection{Galactic Nuclear Abundances} 
\label{sec:galaxyslr} 

For the Milky Way galaxy, the current star formation rate is estimated
to be $\Gamma_{SF} \approx 1.9 \pm 0.40$ $M_\odot$ yr$^{-1}$, where
this result (and error estimate) takes into account many independent
lines of evidence \citep{chompo}.  By definition, the corresponding
supernova rate $\Gamma_{SN}$ is then given by 
\be
\Gamma_{SN} = {\fsuper \over \mbar} \Gamma_{SF} \, . 
\label{snr} 
\ee
For our nominal values $\mbar$ = 0.5 and $\fsuper$ = 0.005, we thus
obtain a supernova rate of 0.019 explosions per year, or, one
supernova every 53 years.

Using the star formation rate discussed above \citep{chompo}, or the
corresponding supernova rate (equation [\ref{snr}]), we can estimate
the abundances of the nuclear species of interest for the Galaxy. The
equilibrium abundance of $^{26}$Al is given by $M[^{26}{\rm Al}]$ =
0.78 $M_\odot$ (1.20 $M_\odot$) for the WW (LC) stellar evolution
calculations, where we haved used $\gamma$ = 1.5 as the index of the
stellar IMF. Similarly, for $^{60}$Fe abundances, we obtain
$M[^{60}{\rm Fe}]$ = 2.77 $M_\odot$ (2.64 $M_\odot$) for the WW (LC)
models. The $^{60}$Fe abundances are nearly the same for the two
models, whereas the LC model predicts $\sim50\%$ more $^{26}$Al; 
these results are a direct reflection of the yields shown in Figure 
\ref{fig:yvgamma}. 

Observations of gamma ray emission, in particular the 1808.65 keV line
from $^{26}$Al, indicate that the current abundance of $^{26}$Al in
the Galaxy is 2.8 $\pm$ 0.8 $M_\odot$ \citep{diehl}. This value is
larger than the estimate found above by a factor of $\sim 2.3$ (where
we use results from the newer LC models). Observations of the
corresponding lines from $^{60}$Fe indicate that the line ratio of
$^{60}$Fe to $^{26}$Al is $R$(Fe/Al) = 0.148 $\pm$ 0.06 \citep{wang}.
In steady state, the ratio of line strengths is proportional to the
ratio of yields expressed in terms of number of nuclei; we thus have
to correct the mass ratio of the yields of $^{60}$Fe to $^{26}$Al by
the ratio 26/60 of their atomic weights. For IMF index $\gamma$ = 1.5,
the ratio of $^{60}$Fe to $^{26}$Al by mass is 0.67 (using LC models)
so that the predicted line ratio becomes $R$(Fe/Al) = 0.29, larger
than the observed line ratio by a factor of about two.  A similar
conclusion was reached by \cite{prantzos}; the line ratio obtained in
that work depends on the maximum stellar mass included in the
analysis, as well as the possible contribution from Wolf-Rayet stars.

Given the masses of $^{26}$Al and $^{60}$Fe inferred for the Galaxy,
we can divide by the total mass in gas (about $10^{10} M_\odot$, e.g.,
\citealt{stahler}) and thereby obtain Galaxy-averaged abundances.  The
resulting mass fractions are $X_{\rm Al} \sim 3\times10^{-10}$ and
$X_{\rm Fe} \sim 10^{-10}$. As expected, these mass fractions are
smaller than those inferred for the early Solar Nebula by an order of
magnitude for both $^{26}$Al and $^{60}$Fe (hence the need for
radioactive enrichment).

To summarize: We find that the predicted values for both the overall
production rate of $^{26}$Al and the ratio of $^{60}$Fe to $^{26}$Al
agree with observations at the factor of two level. This apparent
discrepancy can be interpreted in two ways: We can use this level of
agreement as an estimate for the uncertainties inherent in the results
of the rest of this work (the calculated distributions of nuclear
yields, etc.), i.e., our results would be uncertain by a factor of
two. On the other hand, both the overall abundances and the line ratio
can be understood (to higher precision) if twice as much $^{26}$Al is
produced while keeping the production rate of $^{60}$Fe the same.

Possible uncertainties in the theoretical formulation that affect  
this discrepancy include the following:

[1] The yield of $^{26}$Al calculated here assumes a value of
$\fsuper=0.005$. The IMF for high-mass stars could have a larger
percentage of high mass stars (larger $\fsuper$). However, increasing
$\fsuper$ would increase the yields of both $^{26}$Al and $^{60}$Fe by
the same factor, so that the line ratio would still be in disagreement. 

[2] We have assumed that the index of the IMF $\gamma=1.5$. Smaller
values of the index result in more high-mass stars in the tail of the
distribution and hence larger yields (see Figure \ref{fig:yvgamma}).
However, smaller values of $\gamma$ increase the ratio of $^{60}$Fe to
$^{26}$Al, which works in the wrong direction (see Figure
\ref{fig:massratgam}). If, instead, we increase the value of the index
to $\gamma$ = 2, the mass ratio decreases to 0.46, which implies a
corresponding line ratio $R$(Fe/Al) = 0.20. This value is consistent
(within the error bars) with the observed value of $R$(Fe/Al) = 0.148
$\pm$ 0.06. However, this correction would make the overall abundance
levels smaller (see Figure \ref{fig:yvgamma}), requiring another
adjustment (e.g., larger $\fsuper$) to produce the correct overall
yields.

[3] The star formation rate for the Galaxy could be underestimated.
However, the estimate used here includes multiple constraints (see
\citealt{chompo}), so that changing the star formation rate could
result in disagreement with other observations. In addition, a larger
star formation rate would increase the abundance of $^{26}$Al, but
would leave the line ratio unexplained.

[4] The radioactive yields per star, as determined from the stellar
evolution models, could require modification. Yields from the current
stellar models most likely contain uncertainties at (approximately)
the factor of two level, as indicated by the differences between the
models.

[5] The galaxy could have an additional source of $^{26}$Al. In this
scenario, supernovae would provide all of the $^{60}$Fe and half of
the $^{26}$Al, thereby allowing the star formation rate and stellar
IMF parameters to have their canonical values.  In this case, the
other half must come from other sources, which could include winds
from massive stars \citep{prantzos}, Type Ia supernovae (from white
dwarfs), and spallation sources (including X-winds; see
\citealt{shu2001,gounelle2006}). The presence of spallation sources
are indicated for our own Solar System by evidence for short-lived
$^{10}$Be \citep{mckeegan}, because this SLR cannot be produced via
stellar nucleosynthesis. Spallation from X-winds can also produce the
isotopes $^{36}$Cl, $^{53}$Mn, and $^{41}$Ca at the levels inferred
for the early Solar nebula, but $^{26}$Al is underproduced by factors
of a few (\citealt{shu2001}; see also \citealt{chausgoun,desch} for
additional discussion).

\subsection{Molecular Cloud Nuclear Abundances} 
\label{sec:mcloudslr} 

Next we find the equilibrium abundances of the SLRs on the smaller
scale of molecular clouds using equation (\ref{mequilib}). In this
setting, the star formation rate $\Gamma_{SF}$ is that of the cloud.
Star formation is notoriously inefficient \citep{sal87,mcos}, where
only a small fraction $\sfe$ of the mass of the cloud is turned into
stars during a free-fall time $\taufree$. As a result, the star
formation rate can be written in the form 
\be
\Gamma_{SF} = \sfe {M_{\rm cloud} \over \taufree} \,,
\label{sfrcloud} 
\ee
where $M_{\rm cloud}$ is the mass of the entire cloud. 
The free-fall time $\taufree$ takes the form 
\be
\taufree = \left( {3\pi \over 32 G \rho} \right)^{1/2} \,. 
\label{tffcloud} 
\ee 
The appropriate mass density $\rho$ is that corresponding to number
density $n \approx 10^3$ cm$^{-3}$, so that the free-fall time
$\taufree\approx 1.2$ Myr. If we adopt this value for $\taufree$, then
equation (\ref{sfrcloud}) is exact if we use it as the definition of
the star formation efficiency $\sfe$.  The equilibrium mass fraction
of a given SLR can be determined by combining equations
(\ref{mequilib}) and (\ref{sfrcloud}), i.e., 
\be
X_k = {\sfe \over \ln2} {\halflife\over\taufree} 
{\exstar\over\mbar} \,.
\label{massfrac} 
\ee
Inserting typical values, we obtain 
\be
X_k \approx 4.8 \times 10^{-9} 
\left( {\sfe \over 0.01} \right) 
\left( {\halflife \over 1 \, {\rm Myr}} \right) 
\left( {\exstar \over 0.2 \mu M_\odot} \right) \,.
\label{cloudslr} 
\ee
This result indicates that the mass fraction of $^{26}$Al is expected
to be fall in the range $X_{\rm Al} \approx 3.4-5.3\times10^{-9}$,
where the lower (upper) end of range arises from the WW (LC) stellar
evolution yields (again using $\gamma$ = 1.5 for the stellar IMF).
For both models, the mass fraction of $^{60}$Fe is expected to be 
$X_{\rm Fe} \approx 10^{-8}$. For comparison,
the inferred mass fraction of $^{26}$Al for the Solar Nebula is
$X_{{\rm Al}\odot} \approx 3.8\times10^{-9}$, whereas the mass
fraction of $^{60}$Fe is $X_{{\rm Fe}\odot} \approx 10^{-9}$. For 
both $^{26}$Al and $^{60}$Fe, these mass fractions are an order of 
magnitude larger than those measured for the Galaxy as a whole 
(see Section \ref{sec:galaxyslr}). 

These results have three important implications: [1] For this simple
estimate of the equilibrium abundances of SLRs in molecular clouds,
the mass fractions delivered to star/disk systems are roughly
comparable to those delivered via direction injection (Section
\ref{sec:systemyield}).  [2] The enrichment values are high enough to
account for the estimated abundances of SLRs in the early Solar
Nebula. As a result, a distributed enrichment scenario is viable for
our Solar System \citep{gounelle2008,gounelle2009}. [3] The supernova
yields predict larger mass fractions for $^{60}$Fe, whereas the
meteoritic data from the Solar System indicate higher mass fractions
for $^{26}$Al. This discrepancy, once again, points toward an
additional source for $^{26}$Al (see the discussion of this issue in
Section \ref{sec:galaxyslr}).

This estimate for the distributed contribution to SLR abundances
assumes no losses, i.e., all of the SLRs are captured by the molecular
cloud. In practice, however, supernovae often explode near the cloud
edges, so that some fraction of the SLRs could escape. In addition,
stars near the lower end of the mass range (for supernovae) live for
nearly 30 Myr, thereby allowing them time to leave the clouds before
detonation.  Recent numerical simulations provide some guidance on
this issue. If the supernova explosion has clumpy ejecta, it can be
readily mixed with the surrounding cloud \citep{pan2012};
specifically, the metals from an individual supernova will mix with
$\sim2\times10^4M_\odot$ of cloud material. Since the masses of both
$^{26}$Al and $^{60}$Fe are of order $M_k\sim10^{-5}M_\odot$ (see
Figure \ref{fig:yieldalfe}), the mass fractions for these SLRs are
predicted to be $X_k \sim 10^{-9}$. This result is comparable to, but
somewhat smaller than, the estimate of equation (\ref{cloudslr}).
However, some regions could have higher mass fractions if they are
enriched by multiple supernovae. Simulations of nuclear enrichment for
entire cloud complexes have also been carried out \citep{ake}. These
calculations are also able to reproduce the levels of nuclear
enrichment inferred for our early solar system; however, these
computations are done using a periodic box and do not include losses
from the cloud.

Observations of SLR emission in star forming regions are in their
infancy. A recent review \citep{diehlreview} discusses the current
experimental status for mesaurements of $^{26}$Al lines in Sco-Cen,
Carina, and Orion. Emission from $^{26}$Al is detected in these
regions, but the fluxes (and hence the inferred abundances) are
somewhat lower than expected (given the star formation rate and the
expected nuclear yields). This discrepancy could be interpreted as
evidence for some SLRs being lost from the cloud.  In addition, the
observed emission from Orion originates from a region that is much
larger than the molecular cloud itself; here again, the standard
interpretation is that some fraction of the radioactive material has
escaped from the cloud.  Although a full assessment of the probability
for substantial SLR losses is beyond the scope of this work, it seems
likely that most solar systems will not experience the maximum levels
of enrichment considered here. Nonetheless, distributed enrichment of
SLRs from the background cloud will compete with direct injection into
circumstellar disks.

One important role played by SLRs is their contribution to the
ionization rate. We can illustrate the contribution of distributed
populations of SLRs as follows: The ionization rate $\zeta_k$ due to a
given SLR is given by equation (\ref{zetadef}), although we can
neglect the decaying exponential factor since we are using the
equilibrium abundances. Using the result (\ref{massfrac}) for the mass
fraction, the ionization rate can be written in the form 
\be
\zeta_k = {E_k\over\ionize} {\sfe \over A_k \taufree} 
{\exstar\over\mbar} \,,
\label{ionizeone} 
\ee
where $A_k$ is the atomic number of the nuclear species, $E_k$ is the
energy of the decay products, and $\ionize$ is the energy required for
ionization. Equation (\ref{ionizeone}) specifies the ionization rate 
due to only one SLR; the total ionization rate is determined by a sum 
over all species. Again inserting typical values, we obtain the estimate
\be
\zeta_{SLR} = \sum_k \zeta_k \approx 
3\times10^{-18}{\rm sec}^{-1} \left({\sfe\over0.01}\right) 
\sum_k {1 \over A_k} \left( {E_k\over 1 \, {\rm MeV}} \right) 
\left( {\exstar \over 0.2 \mu M_\odot} \right) \,.
\ee
For typical parameter values, the sum in the above equation is 
about 0.165, so that the benchmark ionization rate due to SLRs 
in molecular clouds $\zeta_{SLR} \approx 5 \times 10^{-19}$ sec$^{-1}$. 
This value is comparable to that expected for SLRs in the early 
Solar Nebula \citep{umenakano,cleevesslr}, but smaller than the 
ionization rate due to cosmic rays in the interstellar medium, 
$\zeta_{CR}\approx1-3\times10^{-17}$ \citep{vandertak}.

\section{Conclusion} 
\label{sec:conclude} 

This paper explores the degree to which young stellar clusters can
influence their constituent solar systems by providing enhanced
abundances of short-lived radionuclides. These SLRs, in turn, affect
disk evolution, disk chemistry, and planet formation by providing
heating and ionization.  Previous work has focused on the possible
enrichment of our own Solar System, and on the total galactic supply
of SLRs. This work generalizes previous treatments by considering SLR
enrichment for typical solar systems residing in a range of cluster
environments.  This section presents a summary of our specific results
(Section \ref{sec:sumresults}) and provides a discussion of their
implications (Section \ref{sec:discuss}).  

\subsection{Summary of Results} 
\label{sec:sumresults} 

Using results from two different sets of stellar evolution
calculations (the WW and LC models), we have calculated the
expectation values of the SLR yields per star $\exstar$, along with
the widths $\sigma_\ast$ of the individual distributions (Table 1).
Although previous work has considered these expectation values, little
focus has been given to the widths of the distributions; these widths
affect the distributions of the radioactive yields provided by
clusters as well as the distributions of SLRs delivered to individual
solar systems. These expectation values for the yields per star have
been calculated as a function of the index $\gamma$ of the stellar IMF
(Figure \ref{fig:yvgamma}), the minimum mass of the progenitor
included in the distribution (Figure \ref{fig:yvmass}), and the time
span included in the treatment (Figure \ref{fig:yvtime}). For both
$^{26}$Al and $^{60}$Fe, the two species of greatest importance, the
expectation values for the SLR mass per star typically fall in the
range $\exstar = 0.1 - 0.3 \mu M_\odot$ (where $1\mu M_\odot$ = 
$10^{-6}M_\odot$). Although the expectation values calculated from the
WW and LC models are roughly comparable, the distributions are 
significantly wider for the LC results (Table 1). 

For clusters with fixed stellar membership size $N$, we have
calculated the distributions of the SLR yields (see Figures
\ref{fig:ldistal} -- \ref{fig:distvmodel}). In the limit of large $N$,
the distributions become nearly gaussian, with expectation values
given semi-analytically via equation (\ref{expectcn}) and with widths
given by equation (\ref{nvariance}).  Clusters with $N=3000$ show
nearly gaussian distributions, whereas somewhat smaller clusters with
$N=1000$ show significant departures from gaussianity. The transition
to the ``large-$N$ limit'' thus takes place near $N \approx 2000$,
about the size of the Orion Nebula Cluster. For systems with $N =
5000$, for example, the typical mass of short-lived radioactive
material provided by the cluster is of order $10^{-3}M_\odot$ = $1 m
M_\odot$, and this value scales linearly with increasing stellar
membership size $N$.

For individual solar systems residing in clusters, we have determined
the distribution of SLR masses provided to their circumstellar disks
(Section \ref{sec:systemyield} and Figure \ref{fig:ss3000}).  For
clusters with large $N$, where the distribution of (total) radioactive
yields per cluster is nearly gaussian, we have derived a semi-analytic
expression for the distribution of SLR masses delivered to solar
systems (see equation [\ref{dpdx}]). This function is in good
agreement with the results found by numerical sampling. We have also
found the corresponding cummulative distribution, which provides the
fraction of solar systems that are exposed to radioactive material
above a given threshold. The fraction of solar systems that could be
enriched at the levels found for the Solar Nebula falls in the range
0.01 -- 0.10.

In addition to finding the distributions (of total cluster yields and
masses delivered to solar systems) for clusters with a given size $N$,
we have also estimated the expectation values integrated over the
entire range of cluster membership sizes (Section \ref{sec:exvalues}).
The ``typical cluster'', in terms of the number of clusters, is quite
small and has a correspondingly small radioactive yield (a few $\mu
M_\odot$). However, the ``typical star'' is predicted to reside in a
larger cluster (which has more stars and more supernovae), so that the
cluster yield that a typical solar system would experience is much
larger, about 0.015 $M_\odot=15 m M_\odot$.  Only a small fraction of
the total mass in SLRs produced by a cluster impinges upon any given
solar system, so the typical mass enrichment is of order
$10pM_\odot=10^{-11}M_\odot$.

We have also considered the connection between the radioactive yields
produced by embedded stellar clusters and the supply of SLRs on larger
scales represented by molecular clouds and the galaxy (Section
\ref{sec:global}). Observations of gamma ray lines indicate that the
galaxy-wide supply of $^{26}$Al and $^{60}$Fe, and their abundance
ratio, is consistent at (only) the factor of two level. More
specifically, the observed line emission from $^{26}$Al is too strong
by a factor of $\sim2$, which could indicate another source (in
addition to supernovae). On the scale of the molecular cloud, we find
that supernovae can enrich the entire cloud at levels comparable to
those inferred for the early Solar Nebula, provided that all of the
SLRs are confined to the cloud. In general, losses of SLRs from the
cloud will reduce the abundances below those of the early Solar
Nebula. Nonetheless, distributed enrichment of SLRs will compete with
the direct injection of SLRs considered in the rest of this paper.

Finally, we have shown that the ionization rate due to SLRs falls in
the range $\zeta_{SLR}\sim1-5\times10^{-19}$ sec$^{-1}$ for both
direct nuclear enrichment in clusters and for distributed enrichment
in molecular clouds. This ionization rate is smaller than the
canonical value usually attributed cosmic rays in the interstellar
medium, $\zeta_{CR}\sim10^{-17}$ sec$^{-1}$. Nonetheless, the CR flux
is often suppressed in young stellar objects \citep{cleevestts}, so
that SLR ionization will be important those systems (see
\citealt{umenakano,cleevesslr}).

\subsection{Discussion and Future Work} 
\label{sec:discuss} 

The mass scales in this problem can be summarized as follows: For the
most abundant isotopes, the yields of SLRs produced by individual
supernovae have masses in the range 10 -- 100 $\mu M_\odot$ (where
$1{\mu}M_\odot=10^{-6}M_\odot$). The corresponding yields per star,
obtained by averaging over the stellar IMF, have masses corresponding
to fractions of $\mu M_\odot$. The yields of SLRs per cluster are much
larger and have masses measured in $m M_\odot$ (where
$1mM_\odot=10^{-3}M_\odot$). Within cluster environments, the typical
mass of SLRs delivered to circumstellar disks is of order $10pM_\odot$
= $10^{-11} M_\odot$, but the range extends up to 100 $pM_\odot$ (but
only for several percent of the solar systems).  For comparison, the
most abundant SLRs found in our own Solar Nebula are thought to have
masses in the range 20 -- 200 $pM_\odot$ (see Table 2).

One implication of this work is that both distributed enrichment in
molecular clouds (Section \ref{sec:mcloudslr}) and direct enrichment
within stellar clusters (Sections \ref{sec:clustyield} and
\ref{sec:systemyield}) can provide significant abundances of SLRs. The
enrichment levels are large enough to contribute to the ionization
rates and --- under favorable circumstances --- large enough to
explain the inferred abundances of SLRs in the early Solar Nebula.
Nonetheless, the manner in which these two enrichment scenarios
compete with each other remains an open question. In both cases, the
amount of radioactive material delivered to a given star/disk system
will be drawn from a wide distribution. For supernova enrichment in
clusters, these distributions are constructed in this paper (see
Section \ref{sec:systemyield}). For distributed enrichment, however,
more work must be carried out to define the distributions. A number of
questions remain, including the amount of nuclear material that is not
seeded into the parental molecular cloud, the amount of cloud material
that is mixed with the supernova ejecta, and the number of high mass
stars that escape the cloud before exploding. All of these quantities 
have a range of values that vary from cloud to cloud, and will 
contribute to the distribution of possible enrichment levels. 

The uncertainties in the nuclear yields from supernovae remain an
important unresolved issue. The level of agreement between the WW and
LC nuclear models (as considered herein) suggest that the yields are
uncertain by a factor of $\sim2$ (see Figure \ref{fig:yieldalfe}).
When averaged over the stellar IMF, the expectation value of the
nuclear yields per star generally differ by somewhat less than a
factor of 2 (Figures \ref{fig:yvgamma} -- \ref{fig:yvtime}). On the
other hand, the differences between the widths of the distributions
are significantly larger (see Table 1), and so are the differences
between the abundance ratios (see Figures \ref{fig:massratgam} and
\ref{fig:massratio}).  This degree of uncertainty is unfortunate, as
the line ratios of emission from $^{60}$Fe and $^{26}$Al are in
apparent disagreement with the stellar nucleosynthesis calculations at
the same factor of $\sim2$ level. More specifically, the observed line
ratios suggest that the Galaxy has a source of $^{26}$Al in addition
to supernovae (which provide the only source of $^{60}$Fe), but the
uncertainties in the nuclear yields are large enough that this
possibility remains inconclusive (see Section \ref{sec:galaxyslr} and
references therein). An important task for the future is thus to
determine the expected nuclear yields with greater specificity.

Orthogonal to the uncertainties in nuclear yields, a number of other
issues should be explored in greater depth. This paper shows that the
expected yields only reach their full values after $\sim30$ Myr, when
the smallest progenitor stars have exploded as supernovae.
Nonetheless, the nuclear yields reach a healthy fraction of their
asymptotic values after $\sim10$ Myr (see Figure \ref{fig:yvtime}).
Since circumstellar disks typically live for $3-10$ Myr \citep{jesus},
or even shorter times \citep{cieza,williams2013}, not every disk will
experience full enrichment. The resulting timing issues are thus
important and should be studied in the future. The direct injection
scenario can suffer from additional inefficiencies due to supernova
fallback, inhomogeneities in the supernova ejecta, and incomplete
capture by the circumstellar disks; all of these issues should be
examined further. With the masses of SLRs that are delivered to solar
systems specified, the implications should also be studied, including
ionization of the gas and heating of planetesimals. These processes,
in turn, will influence disk accretion, planet formation, and the
chemical content of the disk gas. Finally, the long-term properties of
forming planets depend not only on SLRs, but also on radioactive
nuclei with longer half-lives.  The abundances of these isotopes will
be more affected by distributed enrichment, but will be augmented by
direct injection as considered herein.

\medskip
\noindent 
{\bf Acknowledgments:} We thank Ilse Cleeves and Frank Timmes for
useful comments and discussions; we also thank an anonymous referee
for a prompt and constructive report that improved the manuscript.
This work was supported at the University of Michigan through the
Michigan Center for Theoretical Physics and at Xavier University
through the Hauck Foundation.


\end{document}